# Optical Quasi-symmetry Groups for Meron Lattices

Guangfeng Wang [1†], Juan Feng[1†], Tong Zheng[1,2], Yijie Shen[3], Xianfeng Chen[1,4,5*], Bo Wang[1*]

[1]*State Key Laboratory of Photonics and Communications, School of Physics and Astronomy, Shanghai Jiao Tong University; Shanghai, 200240, China.*

[2]*Zhiyuan College, Shanghai Jiao Tong University, Shanghai,200240, China.*

[3]*Centre for Disruptive Photonic Technologies, School of Physical and Mathematical Sciences, Nanyang Technological University, Singapore 637371, Singapore.*

[4]*Shanghai Research Center for Quantum Sciences; Shanghai, 201315, China.*

[5]*Collaborative Innovation Center of Light Manipulations and Applications, Shandong Normal University; Jinan, 250358, China.*

[†]These authors contributed equally to this work.

*Corresponding authors: xfchen@sjtu.edu.cn; wangbo89@sjtu.edu.cn

**Abstract:** We introduce quasi-symmetry groups in optics, emerging from the commutation between mirror operation and the spin–orbit interaction (SOI) of light. Contrary to the principle of symmetry inheritance in free-space optics − where the symmetry of any structured field is strictly constrained by that of its source − we show that strong SOI enables quasi-symmetry-protected formation of meron lattices even when the underlying optical sources violate the nominal rotational symmetry. By analyzing the Hermiticity of the electric-dipole radiation amplitude in a circular polarization basis, we derive an effective mirror operator acting only on a subset of $C_3$ polarized dipole emitters, forming a quasi-symmetry group that commutes with SOI. This quasi-symmetry guarantees exact $C_3$ merons and gives rise to a robust polarization zone within which continuously varying input polarizations generate identical topological textures. Our work establishes quasi-symmetry as a new fundamental principle in optical physics and opens pathways to engineered topological structures of light beyond conventional symmetry constraints.

**Introduction.** Topological quasiparticles such as skyrmions and merons are nontrivial vectoral fields characterized by a robust skyrmion number (*1–5*). Originally proposed in nuclear physics (*6*), skyrmions have been widely investigated in modern magnetic materials, due to their potential in high-density magnetic information storage (*7–9*), and the emergent singular magnetism that leads to nontrivial transport effects such as topological Hall effect (*10, 11*). Merons are regarded as half skyrmion with a half-integer topological number. Due to their logarithmically divergent energy, merons and anti-merons must come into pairs (*12*) to form stably as a lattice of periodic structure as meron lattices (*13*). These structures can be interpreted as the interference pattern of several chiral spin density waves, manifested as a three-component field vector standing wave pattern (*14*). Akin to these magnetic waves, optical analogs of skyrmions have also been reported in various systems, including surface plasmon polaritons (SPPs) (*15–18*), photonic crystals (*19, 20*), and the interference of structured light (*21, 22*), with promising applications found in super-resolution imaging (*23*), optical computing (*24*) and nanoscale metrology (*25*).

To date, despite various optical platforms have been used to demonstrate topological lattices (*18, 26*), none of these cases can violate a symmetry inheritance property of light. The symmetry inheritance from a light source — including all attributes such as amplitude, frequency, momentum and polarization — is ubiquitously maintained during its free-space propagation, serving as a fundamental property throughout all linear optical domains and even general wave systems (*27–30*). This property essentially governs the construction of topological lattices through interference or superposition from several waves (*15, 31–35*). For instance, it was demonstrated that only optical sources with fourfold or sixfold rotational symmetry are permitted to form skyrmion or meron lattices with identical symmetries (*15, 31*), while topological quasicrystals from higher-rotational symmetry systems also exhibit strict rotational symmetry inheritance (*35–37*).

Very recently, the concept of *quasi-symmetry* has been introduced as an extension to the conventional group theory describing the degeneracy of bands, enabling the determination of energy splitting effects induced by symmetry-lowering perturbations in condensed matter physics (*38–40*). Unlike the strict crystalline symmetry, *quasi-symmetry* is an approximated symmetry which commutes with the lower-order $k \cdot p$ Hamiltonian in the presence of spin-orbit interaction (SOI), leading to unexpected topology and new classifications of materials beyond the constraints of usual spatial symmetries. For instance, a *quasi-symmetry* emerges when a mirror operation acts on a subset of an object, in contrast to exact symmetry that uniformly acts on the whole object (*39*). Although the SOI of light, namely, the interplay between optical spin and orbital angular momentum, has been widely explored (*41–43*), a *quasi-symmetry* group arising from the nontrivial interplay between symmetry and SOI has not yet been reported in any optical regimes.

In this work, we demonstrate the very first quasi-symmetry-protected optical meron lattices in free space due to strong SOI of light. The quasi-symmetry arises from the Hermitian conjugation of the scattering amplitude of electric dipole field in a circular polarization basis. Physically, this Hermiticity leads to a particular mirror operation of dipole's polarization upon a subset of $C_3$ polarized dipole system. In contrast to previous understandings of optical symmetry inheritance, our study shows that quasi-symmetry groups – which obviously violate $C_3$ symmetry – can still generate perfect $C_3$ merons due to strong SOI of light. Moreover, the probability of polarization states that can generate merons rises abnormally as the SOI enhances, and it ultimately results in a robust polarization zone, within which all continually changed polarization states can lead to $C_3$ merons without symmetry constraints. We generalized this concept in higher-order rotational symmetries $C_{2N+1}$ and general SOIs beyond dipole radiation, revealing the universal nature of *quasi-symmetry-protected topology* in nonparaxial optical fields.

**The definition of optical quasi-symmetry group and its impact on topological optical lattices.** The main concepts of this work are presented in Fig.1. We consider a set of $N = 3$ linear-polarized monochromatic dipoles $\mathbf{E}_i$ at the vertices of an equilateral triangle in the *xy* plane. The distance between a dipole and the origin is $r$, and the azimuthal angles of the dipoles are $\varphi_i = 2\pi(i-1)/N$, $i = \{1, ..., N\}$. Each dipole features an in-plane polarization angle $\Phi_i = \varphi_i + \xi_i$, with $\xi_i \in [0, \pi]$ being the angle between a dipole's polarization and its azimuthal vectors $\mathbf{r}_i = r(cos\varphi_i, sin\varphi_i)$. This notation allows us to conveniently compare a dipole system to a $C_3$ symmetric system. For $\xi_i = \xi$, the three-dipole system possesses a threefold rotational symmetry about *z*-axis, since $\hat{\mathcal{R}}_{2\pi/3}\Psi = \Psi$. Here, $\hat{\mathcal{R}}_{2\pi/3}$ is the three-fold rotation operator, and $\Psi = \sum_{i=1}^{3} \delta(\mathbf{r} - \mathbf{r}_i)\mathbf{E}_i$ denotes the electric field of the three-dipole system on the $z = 0$ plane, as shown in Fig.1A. As the dipoles propagate in free space, an ideal Stokes meron is generated at distant *xy* planes, where the Stokes parameters ($S_1$, $S_2$, $S_3$) are formulated by the superposition of the *xy*-plane electric field components (Fig. 1C). This is a natural result of optical symmetry inheritance: perfect topological textures with *N*-fold rotational symmetry are produced under interference sources with a global $C_N$ symmetry (*15, 29, 31, 36, 37*). Crucially, owing to the odd parity of the third Stokes parameter $S_3(x, y)$, non-zero merons can only be formed when *N* is odd (Section S2 for details); this leaves *N* = 3 the only case to generate periodic meron lattices. Instead of studying this trivial case, we raise a question to seek the possibilities of generating perfect $C_3$ merons when $\xi_i$ are arbitrarily chosen such that $\hat{\mathcal{R}}_{2\pi/3}\Psi \neq \Psi$.

To answer this question, we must consider the intrinsic SOI of the dipole field during its propagation in free space. For a dipole source $\mathbf{E}_2$ (the purple arrow in Fig.1B), its field distribution in the momentum space is $\mathbf{E}(\mathbf{k}) \cong \mathbf{E}_2 - \mathbf{k}(\mathbf{k} \cdot \mathbf{E}_2)$, where $\mathbf{k}$ is the unit vector of the observation point defined by the spherical coordinates $(\theta, \varphi)$. Here, $\theta$ and $\varphi$ denote the polar and azimuthal angle of the radiated wave, respectively (Fig.1D). $\mathbf{E}(\mathbf{k})$ is manifested as the rotation of $\mathbf{E}_2$ on a conical $\mathbf{k}$-distribution waves with a fixed deflection angle $\theta$, as shown in the left panel of Fig.1D. This form of spherical projection of the incident polarization $\mathbf{E}_2$ onto the $\mathbf{k}$ sphere, namely, $\mathbf{k}(\mathbf{k} \cdot \mathbf{E}_2)$, naturally sheds light on an analogy of $k \cdot p$ Hamiltonian associated with SOI (*39*). To better quantify the SOI in dipole radiation, we rewrite $\mathbf{E}_2 - \mathbf{k}(\mathbf{k} \cdot \mathbf{E}_2)$ as $\hat{\mathcal{H}}(\mathbf{k})\mathbf{E}_2^C$. Here, $\mathbf{E}_2^C = (e^{-i\Phi_1}, e^{i\Phi_1})$ is $\mathbf{E}_2$ formulated in the circular polarization basis, and $\hat{\mathcal{H}}(\mathbf{k})$ is an SOI operator reduced in *xy* dimension (Section S1 for details):

$$\widehat{\mathcal{H}}(\mathbf{k}) = \begin{pmatrix} A & -Be^{-2i\varphi} \\ -Be^{2i\varphi} & A \end{pmatrix} \quad (1)$$

In Eq. (1), $A = 1 + \cos^2\theta$ and $B = \sin^2\theta$, represent the spin-maintained and spin-flipped component, respectively (*44*). Appling Eq. (1) to a linearly polarized dipole source with a polarization angle $\Phi = \varphi + \xi$, the resulting radiation field in the circular basis is:

$$\mathbf{E}(\mathbf{k}) \propto \begin{pmatrix} Ae^{-i(\varphi+\xi)} - Be^{-i(\varphi-\xi)} \\ Ae^{+i(\varphi+\xi)} - Be^{+i(\varphi-\xi)} \end{pmatrix} = \begin{pmatrix} \mathcal{S}e^{-i\varphi} \\ \mathcal{S}^* e^{+i\varphi} \end{pmatrix} \quad (2)$$

Here, we have defined $\mathcal{S}(\xi) = Ae^{-i\xi} - Be^{+i\xi}$. Therefore, the contribution of a dipole's radiation field is separated to a location-originated part "$e^{\pm i\varphi}$" and a polarization-originated part "$\mathcal{S}$", where $\mathcal{S}$ is mediated by the source's polarization "$\xi$" and the strength of SOI. We have noticed that the $\mathcal{S}$ parameter has a conjugate property: $\mathcal{S}^*(\xi) = -\mathcal{S}(\pi - \xi)$, which bridges two polarization states: $\Phi = \varphi + \xi$ and $\Phi = \varphi + \pi - \xi$. Physically, converting a dipole's polarization from $\varphi + \xi$ to $\varphi + \pi - \xi$ is achieved by introducing an in-plane mirror operator $\widehat{\mathcal{M}} = -\widehat{V}^\dagger \widehat{\mathcal{R}}_\varphi \widehat{\sigma}_z \widehat{\mathcal{R}}_{-\varphi} \widehat{V}$ that acts on the polarization of dipole $\mathbf{E}_2$ along an axis perpendicular to $\mathbf{k}_\varphi$. Here, $\widehat{\sigma}_z$ is the third Pauli matrix, $\widehat{V}$ denotes the unitary transformation between linear and circular bases (Section S1 for details), and the mirror axis is the dashed line in the right panel of Fig.1D. Accordingly, the radiation field $\mathbf{E}(\mathbf{k})$ along $\mathbf{k}_\varphi$ is also mirror flipped (Fig.1D, right panel), namely, $\widehat{\mathcal{M}}\widehat{\mathcal{H}}(\mathbf{k}_\varphi)\mathbf{E}_2 = \widehat{\mathcal{H}}(\mathbf{k}_\varphi)\widehat{\mathcal{M}}\mathbf{E}_2$. This leads to a commutation relation between $\widehat{\mathcal{M}}$ and $\widehat{\mathcal{H}}(\mathbf{k}_\varphi)$:

$$[\widehat{\mathcal{M}}, \widehat{\mathcal{H}}(\mathbf{k}_\varphi)] = \widehat{\mathcal{M}}\widehat{\mathcal{H}}(\mathbf{k}_\varphi) - \widehat{\mathcal{H}}(\mathbf{k}_\varphi)\widehat{\mathcal{M}} = 0. \quad (3)$$

Equation (3) is the main result of this work, which brings forth the definition of a mirror operator $\widehat{\mathcal{M}}$ associated with the direction $\mathbf{k}_\varphi$. This commutation relation is analogical to that found in a semi-metal, in which a partial mirror operation also commutes with the lower-order *k·p* Hamiltonian in the presence of SOI (*39*). Likewise, Eq. (3) leads to new classifications of symmetry groups beyond the constraints of usual spatial symmetries. Notably, by observing $\mathbf{E}(\mathbf{k})$ from increased $\theta$, the in-plane components of radiation fields generated by $\mathbf{E}_2$ and $\widehat{\mathcal{M}}\mathbf{E}_2$ are gradually aligned with their mirror axis, and they are eventually overlapped under the grazing angle limit ($\theta \to \pi/2$) with the maximized SOI, as shown in Fig.1E. This means that, as SOI intensifies, the in-plane fields from arbitrary polarized dipole sources and their mirror counterparts become indistinguishable, and the angular location of dipole $\varphi$ solely determines the polarization, and $\mathbf{E}(\mathbf{k}) \perp \mathbf{k}_\varphi$. Note that the direction of $\mathbf{k}_\varphi$ is fixed for a dipole when $N$-dipole composes a system with $\varphi_i = 2\pi(i-1)/N$. In this case, by incorporating the mirror operator $\widehat{\mathcal{M}}$ into a $C_N$-dipole system and partially act on any subsets of the dipoles—defined as a $\widehat{\mathcal{M}}_{\text{eff}}$ operator (Fig.1B)—we find the corresponding $\Psi_q = \widehat{\mathcal{M}}_{\text{eff}}\Psi$ belong to optical *quasi-symmetry* $C_N$ groups (q-$C_N$) which are able to produce perfect topological lattices without polarization constraints.

In Fig. 1F, we summarized all symmetry groups for generating perfect $C_3$ merons under varying SOIs mediated by the radiation angle $\theta$. The cube is a polarization space determined by the relative polarization angle $\xi_i$, with $\xi_i \in [0, \pi]$, denoting all possible linear polarization combinations of a three-dipole system depicted in Fig. 1A (The sphere is presented for better visualization). For weak SOI, the generation of merons are symmetry-guaranteed, with four symmetry groups: one is $C_3$ and the others are anti-$C_3$ (will be discussed in the next section). As the SOI increases, it destroys the anti-$C_3$ group, but $C_3$ group is survived. For strong SOI, the q-$C_3$ arises as three additional branches to dominate the produce of merons. Notably, the $C_3$ and q-$C_3$ will merge as a bulky polarization zone, which contains continually changed polarization states to form perfect $C_3$ merons, indicating a strong polarization robustness to induce topological lattices of light by strong SOI.

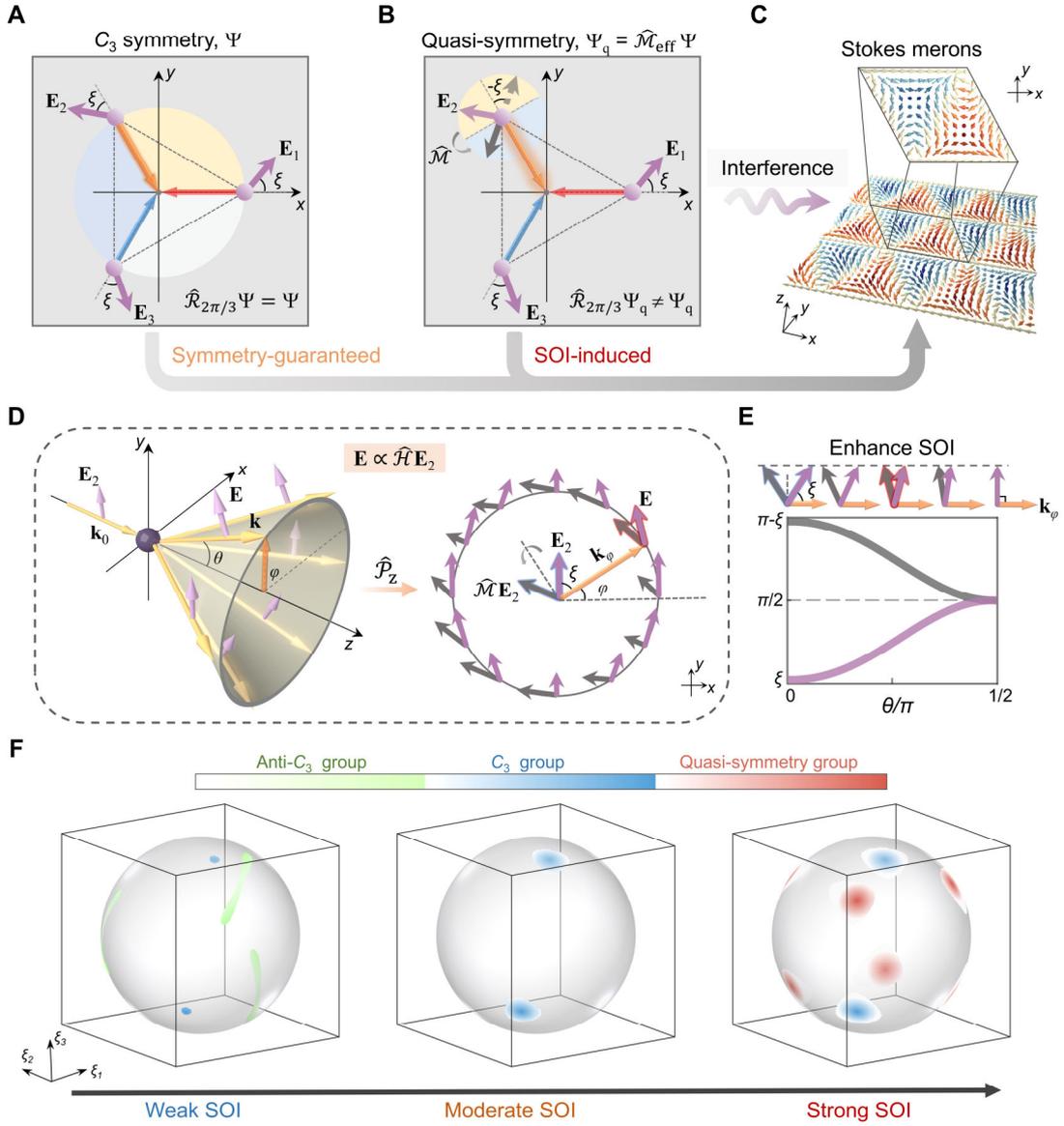

**Fig.1. Concept of the optical *quasi-symmetry* group.** (**A**) A $C_3$-symmetric dipole system $\Psi$ in the $z = 0$ plane. (**B**) Definition of a Quasi-symmetry $C_3$ system. $\Psi_q$ is formed by applying a partial mirror operator $\widehat{\mathcal{M}}_{\text{eff}}$ to $\Psi$ in (**A**), which implements a mirror operator $\widehat{\mathcal{M}}$ on a dipole along the axes perpendicular to **r** (the dashed border). (**C**) Stokes meron lattices (inset shows a unitcell) generated by the interference of the three dipoles at a distant observation plane $z$. This topological lattice is either symmetry-guaranteed by the $C_3$ system or SOI-induced by the q-$C_3$ system. The 3D arrows represent the orientation distribution of the Stokes vectors $\mathbf{S}(x, y) = [S_1, S_2, S_3]$ and the color indicates $S_3$ "up" (Red) or "down" (Blue). (**D**) Left panel: Radiation field from a dipole $\mathbf{E}_2$ produces a conical **k**-distribution (light yellow surface and arrows) and associated adiabatically rotating field $\mathbf{E}(\mathbf{k})$ (purple arrows). Right panel: Projected

in-plane field distribution $\mathbf{E}(\mathbf{k})$ at a fixed deflection angle $\theta$ originating from $\mathbf{E}_2$ and its mirror image $\widehat{\mathcal{M}}\mathbf{E}_2$. $\widehat{\mathcal{P}}_z$ = diag (1, 1, 0). (**E**) Evolution of the relative angle between the projected mirror polarization pair and the $\mathbf{k}_\varphi$ (orange arrows) direction as $\theta$ increases from 0 to $\pi/2$. Initial polarization angle is $\xi \in [0, \pi]$. (**F**) Typical polarization symmetry groups for ideal merons in the polarization cube $(\xi_1, \xi_2, \xi_3)$ with different SOIs. Distribution of states on the spherical surface corresponding to distinct symmetry groups: anti-$C_3$ (Green), $C_3$(Blue), and q-$C_3$ (Red). Color gradient from dark to light indicates the deviation degree from exact symmetry operations, quantified by Euclidean distance in parameter space.

**Polarization symmetry groups for $C_3$ merons without SOI.** We first examine the polarization configurations that produce merons under paraxial approximation in the absence of SOI ($\theta \rightarrow 0$). The cube in Fig. 2A contains all linear polarization states labeled as $(\xi_1, \xi_2, \xi_3)$, wherein each colored site indicates a polarization state leading to a perfect lattice. These sites are identified by comparing the formed Stokes lattice from $\widehat{\mathcal{H}}(\mathbf{k})\Psi$ with a standard $C_3$ meron, with the normalized overlapped factor $\mathcal{K}(\xi_1, \xi_2, \xi_3) \rightarrow 1$(Section S4 for details). Fig.2A shows four groups in the polarization cube, wherein the diagonal group is the known $C_3$ group (Type I), as one polarization state illustrated in the left panel of Fig. 2B. The remaining three groups are labeled as Type II to IV, forming three lines in parallel to that of Type I. We find that Type II to IV belong to the same symmetry group merely with a different initial polarization difference. We term to them as an "anti-$C_3$" group, because for each state, the polarization and location of dipoles follow opposite rotation operations: $\widehat{\mathcal{R}}_{2\pi/3}\mathbf{r}_i=\mathbf{r}_{i+1}$, and, $\widehat{\mathcal{R}}_{2\pi/3}^{-1}\mathbf{E}_i=\mathbf{E}_{i+1}$ for $i = 1,2,3$(cyclic). This additional group exhibits a special "anti-$C_3$" vortex configuration characterized by reversed cyclic sequence relative to $\mathbf{r}_i$, distinct from the sequential order of $C_3$ group. (Table S1 for details).

The normalized Stokes vectors $\mathbf{s}(x,y) = [S_1, S_2, S_3]/S_0$ corresponding to Type I and Type II are shown in Fig.2C, exhibiting triangular meron textures quantified by the topological number $N_{sk} = (1/4\pi) \iint_A \mathbf{s} \cdot (\partial_x \mathbf{s} \times \partial_y \mathbf{s}) \, dxdy \simeq \pm 0.5$ ($A$ denotes the region of a unitcell of lattices). As demonstrated in Fig.2D, the merons are mapped onto the upper or lower hemisphere of the Poincaré sphere, where the boundary ($S_3$= 0) corresponding to the equator (red circles), and the center of triangular lattices with vertically directed $\mathbf{s} = \pm\mathbf{z}$ correspond to the circular points. It is noteworthy that the $C_3$ group of polarization states exclusively produces anti-type merons with a vorticity $m = -1$, while the other anti-$C_3$ group enables the formation of various topological textures with $m = +1$ spanning Néel-type, intermediate-type (twisted textures) and the Bloch-type (*18*) , due to the helicity $\gamma$ being continuously modulated by $\xi$ varying from $-\pi$ to $\pi$ (Section S3 for details).

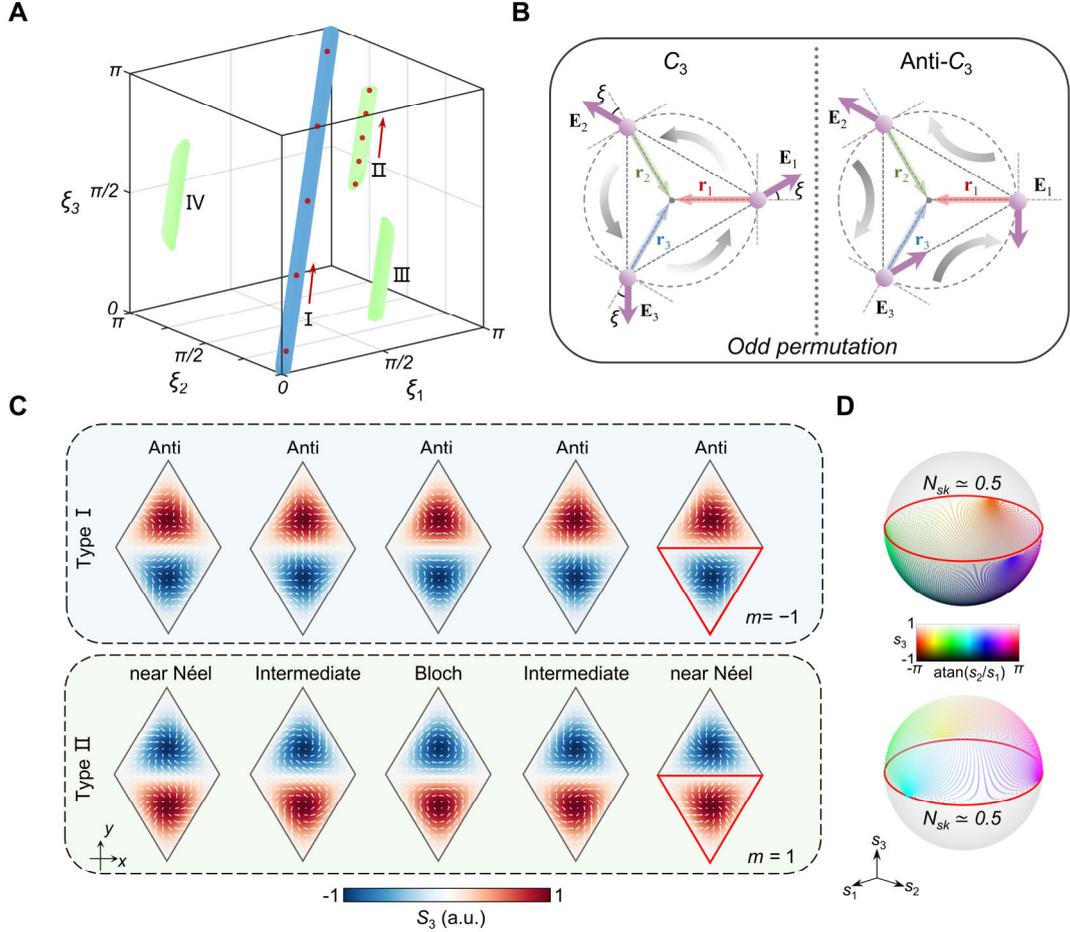

**Fig.2. Polarization symmetric configurations for ideal Stokes meron lattices in the absence of SOI.** (**A**) Polarization states to form $C_3$ merons in the polarization cube in the paraxial regime ($\theta \to 0$). Type I (blue solid line) corresponds to $C_3$ polarization group, while Types II-IV (green solid lines) represent anti-$C_3$ groups. (**B**) Schematic of a $C_3$ (left panel) and anti-$C_3$ (right panel) polarization state. For the $C_3$ state, $\hat{\mathcal{R}}_{2\pi/3}\mathbf{r}_i=\mathbf{r}_{i+1}$, and $\hat{\mathcal{R}}_{2\pi/3}\mathbf{E}_i=\mathbf{E}_{i+1}$(left panel, curved arrows). For the anti-$C_3$ state, $\hat{\mathcal{R}}_{2\pi/3}\mathbf{r}_i=\mathbf{r}_{i+1}$, and, $\hat{\mathcal{R}}_{2\pi/3}^{-1}\mathbf{E}_i=\mathbf{E}_{i+1}$ (right panel, curved arrows). $i = 1,2,3$ (cyclic order)

(**C**) Stokes vectors **S**(*x,y*) over triangular unitcells corresponding to the selected polarization states (red points) in (A). These meron pairs exhibit different vorticity $m$ and helicity $\gamma$. The white arrows represent the in-plane vectors ($S_1$, $S_2$) and color indicates the $S_3$ component. (**D**) The normalized Stokes vectors in the triangular unitcells (the last panels in (C)) represented by brightness ($s_3$) and colormap (atan($s_2/s_1$)), are mapped onto the Poincaré spheres with topological number $N_{sk} = \pm 1/2$. The meron boundaries correspond to $S_3 = 0$ (red circles).

**SOI-driven dynamics of the polarization symmetry groups.** A clear evolution of polarization configurations for $C_3$ merons is observed in the 3D parameter space with varying SOI. Fig.3A to D present several typical states across different SOI strengths as quantified by the radiation angle $\theta$. It can be seen that as $\theta$ increases from 0 to $0.1\pi$, corresponding to a weak SOI, the anti-$C_3$ groups in the paraxial regime are distorted, deviating from their original spatial regions (Fig.3A). This means that the anti-$C_3$ group is fragile from the SOI due to its lack of global $C_3$ symmetry (Section S8 for details). Meanwhile, the diagonal line corresponding to the $C_3$ group remains unchanged, as it was guaranteed by the global symmetry. In the moderate SOI regime ($\theta = 0.25\pi$), the anti-$C_3$ groups are completely eliminated, while the $C_3$ region expands slightly around its center (Fig. 3B). When the SOI is further enhanced (Fig. 3C and D), the allowed number of polarization states increases dramatically, and it eventually forms a bulky star-like region, which is dominated by four diagonal branches as shown in Fig.3D. In the bottom panel of Fig.3, we selected several cut-planes to explicitly present the dynamics of different symmetric groups with varying SOI. These cut-planes clearly show the preservation of $C_3$, the elimination of anti-$C_3$, and the proliferation of q-$C_3$ groups with enhanced SOI. Particularly, for strong SOI, the possibility to observe meron lattices is significantly increased due to q-$C_3$ groups. The continuous evolution of polarization combinations for perfect lattices is shown in the supplementary movies.

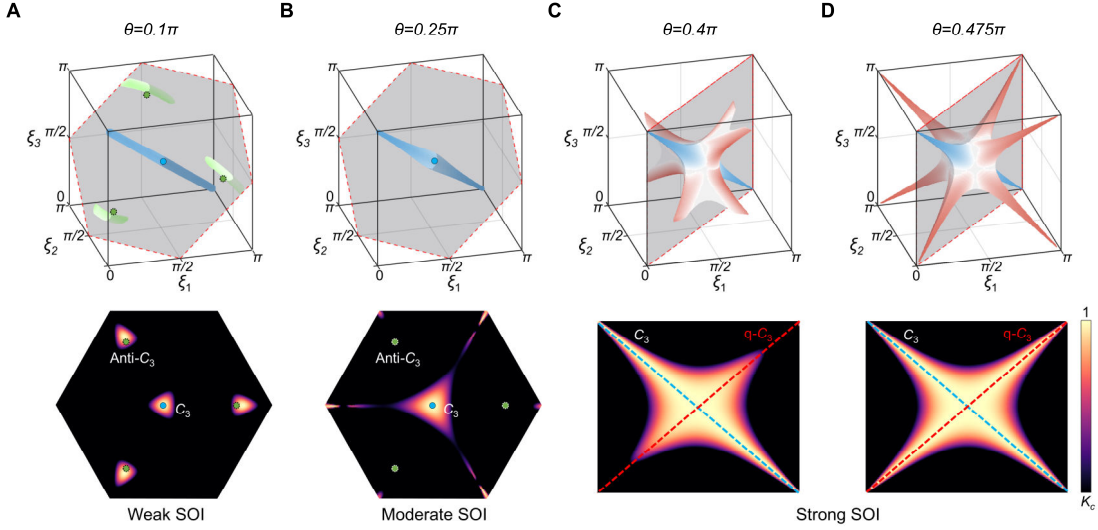

**Fig.3. Evolution of polarization symmetry groups with varying SOIs. (A)—(D)** Polarization states in the cube to form perfect $C_3$ merons across distinct SOI regimes: weak ($\theta = 0.1\pi$), moderate ($\theta = 0.25\pi$), and strong ($\theta = 0.4\pi$ and $0.475\pi$). Polarization groups are represented by various colors using the same scheme as Fig.1F. The hexagonal and rectangular cut-planes are plotted for better symmetry group characterization, as solutions shown in the bottom panels of this figure. The threshold $K_c$ is set to 0.99.

**The proliferation of quasi-symmetry group under strong SOI.** We focus on the evolution of the q-$C_3$ group under different SOIs. Based on the formalism of Eq. (2), we give the expression of the $\mathcal{K}$ factor associated with a q-$C_3$ polarization state $(-\xi, \xi, \xi)$ (Section S4 for details):

$$\mathcal{K}(-\xi, \xi, \xi) = \frac{\sum_{i \neq j}^{3} |\sin(\varphi_{ji} + \beta_{ij})|}{\sqrt{3}\sqrt{\sum_{i \neq j}^{3} |\sin(\varphi_{ji} + \beta_{ij})|^2}}$$

Here, $\varphi_{ji} = 2\pi(j-i)/3$, $\beta_{ij} = \arg(\mathcal{S}_i \mathcal{S}_j^*)$ characterizes the phase difference of the radiation fields between any two dipoles. For the two dipoles with the same polarization $\xi$, $\beta_{ij} = \beta(\xi, \xi) = \arg(|\mathcal{S}(\xi)|^2) = 0$. Therefore, a $C_3$ dipole system naturally has $\mathcal{K}(\xi, \xi, \xi) = 1$ to produce perfect merons. In the q-$C_3$ system, $\beta_{ij} = \beta(-\xi, \xi; \theta)$, which possesses an odd parity with respect to $\xi = \pi/2$, as shown in the $(\theta, \xi)$ space (Fig. 4A). For small $\theta$, perfect merons are only generated around $\xi = \pi/2$. In this case, q-$C_3$ and $C_3$ are degenerate and indistinguishable. As $\theta$ increases, $\beta(-\xi, \xi; \theta)$ rapidly converges to 0 for any $\xi \in [0, \pi)$, consequently driving the $\mathcal{K}$ factor to approach 1 (Fig.4B). As shown in Fig.4D, we selected several Stokes vectors **S**(*x,y*) from the same $(-\xi, \xi, \xi)$ dipole system but with different $\theta$. These lattices present a progression of merons from broken symmetry to "perfect" $C_3$ symmetry as the SOI enhances. Besides, we calculated the density of polarization states $\rho(\mathcal{K} > K_c)$ for generating merons. Here, $\rho(\mathcal{K} > K_c)$ stands for the percentage of polarization states among all q-$C_3$ configurations to generate perfect merons with a criterion $\mathcal{K} > K_c$. As shown in Fig. 4C, $\rho(\mathcal{K} > K_c)$ grows rapidly as $\theta$ enters the strong SOI region, with the peak occurs at $\theta = \pi/2$, which corresponds to a critical SOI condition: $A = B$. Mathematically, this critical SOI condition is guaranteed by the Hermitian conjugation of dipole's radiation amplitude $\mathcal{S}(\xi) = \mathcal{S}^*(-\xi)$, consistent with the commutation relation Eq. (3). The fact that the critical SOI condition occurs at the grazing limit ($\theta \to \pi/2$) is also originated from the radiation nature of electric dipole. For general multipole scattering systems such as electric quadrupoles, magnetic dipoles, etc., *A* and *B* can even be complex values, and the critical SOI condition will shift to $\theta < \pi/2$. In general, a critical SOI condition to reach maximal $\rho(\mathcal{K} > K_c)$ from the quasi-symmetry groups is:

$$\begin{cases} |A| = |B| \neq 0 \\ \phi_{BA} = \arg(B) - \arg(A) = m\pi, m \in \mathbb{Z} \end{cases} \exists \, \theta \in (0, \pi/2]$$

This condition has been verified with customized Mie scattering and focused beam systems to obtain optimized $\theta$ and different $\widehat{\mathcal{H}}(\mathbf{k})$ (Section S6 for details).

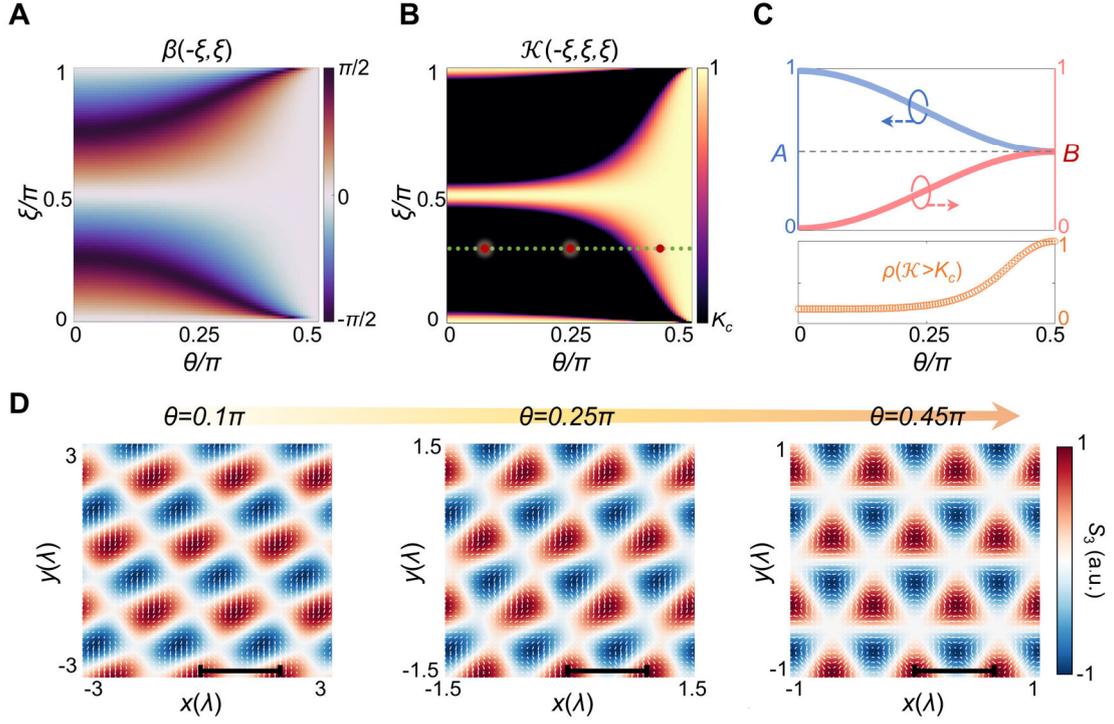

**Fig.4. Details about the proliferation of q-$C_3$ symmetry group.** Calculated $\beta$ (**A**) and (**B**) $\mathcal{K}$ factor with different diffraction angle $\theta$. (**C**) Calculated SOI coefficients *A* (spin-maintained part, blue curve) and *B* (spin-flipped part, red curve) as functions of $\theta$, along with the evolution of the normalized density of states $\rho(\mathcal{K} > K_c)$. (**D**) Stokes vectors constructed from a q-$C_3$ dipole system $(-\xi, \xi, \xi)$ crossing different $\theta$ as marked in (B). Here, $K_c = 0.99$ and $\xi = \pi/3$ (green dash line in (B)). Scale bars: $2\lambda/(3\sin\theta)$.

**Discussion**

In conclusion, we have discovered optical quasi-symmetry groups in a $C_N$ system composed of *N* optical dipole sources. This quasi-symmetry is defined by one or a few mirror operations on a subset of the system, which is originated from the commutation relation of our Eq. (3). Moreover, we further extend this concept to arbitrary elliptically polarized dipoles systems (Section S5 for details). Due to the strong SOI of light in free space, quasi-symmetry groups can produce perfect meron lattices, deepening our understanding of previously symmetry-guaranteed conditions to generate topological lattices of light.

Since an odd number $N$ is required to form the Stokes nontrivial textures, this quasi-symmetry property does not exist in other crystalline symmetric groups for even $N$ (Section S2 for details). Therefore, we extended the dipole systems to $N$ = 5, 7 and 9, forming distinct quasi-crystalline Stokes textures. Akin to the operation in $C_3$, we introduce similar operators $\widehat{\mathcal{M}}_{\text{eff}}$ to these highly rotational symmetry $C_N$ dipole sources to construct q-$C_N$ groups, and presented $C_N$ Stokes textures under a strong SOI, as shown in Fig.5A. Ideal quasi-crystalline intensity and spin textures strictly possessing $N$-fold rotational symmetry are generated due to the introduce of *quasi-symmetry* mirror operation, demonstrating the generality of our *quasi-symmetry* mechanism (Fig.5B-C). Particularly, the formation of quasi-periodic textures reveals unprecedented topological structures of light (Fig. 5D), which exist not only in SPP systems (*35–37*) but also in nonparaxial light in free space.

Owing to the ubiquitous nature of SOI, *quasi-symmetry* may be discovered in diverse nontrivial vector fields, including spin angular momentum (*16*), energy flux (*45, 46*) and pseudospin (*19*) in real or momentum space, even spatiotemporal fields (*23*). Moreover, analogous phenomena about approximate symmetry may also be revealed in other wave systems including water wave (*32, 33*), acoustic wave (*47*).etc., where effective SOI can be constructed similarly. Our work leads to extended optical symmetry groups for the observation and manipulation topological structures of light, possible useful for applications in exotic light-matter interactions (*48*) such as constructing topological optical lattices to manipulate atoms (*49*) and nanoparticles (*50*), or generate higher order harmonics (*51, 52*).

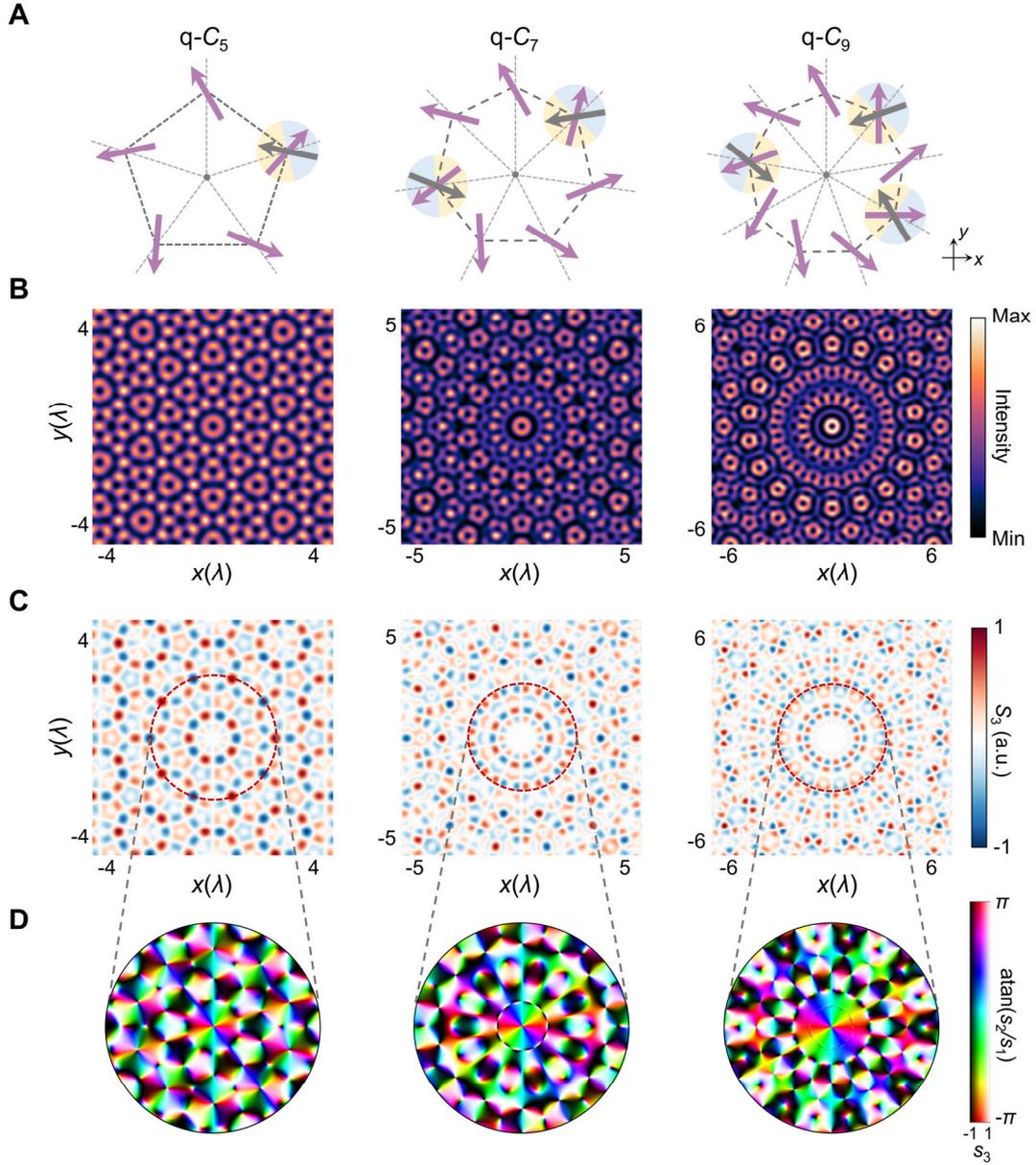

**Fig.5. Quasicrystal-like topological textures in *quasi-symmetry* polarization groups under strong SOI.** (**A**) Schematic of *N*-fold rotational dipole sources with *N* = 5, 7, and 9. Random partial mirror operations (gray arrows) are embedded into these $C_N$ sources (purple arrows) where the mirror axes are strictly perpendicular to the corresponding $\mathbf{r}_i$ of that dipole. Initial angles of these dipoles are set as $\xi = \pi/6$. (**B**) Quasicrystal-like $C_N$ light intensities and (**C**) $S_3$ distributions on the observation plane with a strong SOI at $\theta = 0.45\pi$. (**D**) Selected Stokes vector textures marked by the red circle in (C) exhibiting mixed topological quasiparticles including vortices, skyrmions and meron pairs.

**Acknowledgments**

**Funding:** This work was supported by National Science Foundation of China (12274296 and 12192252) and Shanghai Jiao Tong University 2030 Initiative. B.W. is sponsored by Yangyang Development Fund.

**Author contributions:** B.W conceived the idea. B.W and X.C supervised this work. G.W systematically developed the theory and conducted the analysis. J. F performed early-stage simulations. B.W and G.W wrote the manuscript and prepared the supplementary materials. Y.S and T.Z assisted in manuscript revision. All the authors discussed the results. G.W and J. F contributed equally to this work.

**Competing interests:** The authors declare that they have no competing interests.

**Data and materials availability:** All data needed to evaluate the conclusions in the paper are present in the paper and/or the Supplementary Materials.

# Supplementary Materials for

## Optical Quasi-symmetry Groups for Meron Lattices

Guangfeng Wang *et al.*

Corresponding author: Xianfeng Chen, xfchen@sjtu.edu.cn;

Bo Wang, wangbo89@sjtu.edu.cn

**This PDF file includes:**

Supplementary Text
Figs. S1 to S18

Table S1
References

# Section S1: Matrix representation of the SOI operator in dipole scattering

To describe the SOI-mediated behaviors of dipole fields, we derived the transformation matrix of the SOI operator in the circular polarization basis. We employ the basis of circular polarization relative to the $z$-axis due to the azimuthal symmetry of the optical system. Considering the complex amplitude of electric field $\mathbf{E}_\mathrm{L} = (E_x, E_y, E_z)^T$ in the global Cartesian frame with the linear polarization basis $(\mathbf{e}_x, \mathbf{e}_y, \mathbf{e}_z)$, the basic vectors and electric field amplitudes in the circular polarization basis are written as:

$$\mathbf{e}_\pm = \frac{\mathbf{e}_x \pm i\mathbf{e}_y}{\sqrt{2}}, E_\pm = \frac{E_x \mp iE_y}{\sqrt{2}} \tag{S1}$$

Thus, $\mathbf{E}_\mathrm{C} = (E_+, E_-, E_z)^T$ in basis $(\mathbf{e}_+, \mathbf{e}_-, \mathbf{e}_z)$, where the transformation is realized via a unitary matrix $\hat{V}$, that is:

$$\mathbf{E}_\mathrm{L} = \hat{V}\mathbf{E}_\mathrm{C}, \hat{V} = \frac{1}{\sqrt{2}}\begin{pmatrix} 1 & 1 & 0 \\ i & -i & 0 \\ 0 & 0 & \sqrt{2} \end{pmatrix} \tag{S2}$$

All following analyses adopt the circular basis $(\mathbf{e}_+, \mathbf{e}_-, \mathbf{e}_z)$, with the subscript C omitted for conciseness. We further consider the far-field approximation of the dipole, associating the radiation momentum with spherical coordinates $\mathbf{k} \propto (\sin\theta\cos\varphi, \sin\theta\sin\varphi, \cos\theta)$. Here, $\theta \in (0, \pi/2]$ is the polar angle and $\varphi \in [0, 2\pi)$ is the azimuthal angle relative to the $z$-axis. The form of scattered field in momentum space is $\mathbf{E}(\mathbf{k}) \propto -\mathbf{k} \times (\mathbf{k} \times \mathbf{E}_0)$, where the double cross-product operator can be reformulated as a series of purely 3D rotations of the coordinate frame accompanied by the rotations of the local polarization states (ensure the orthogonality $\mathbf{k} \cdot \mathbf{E} = 0$), that is $-\mathbf{k} \times (\mathbf{k} \times \cdots) \propto \hat{\mathcal{R}}_z(-\varphi)\hat{\mathcal{R}}_y(-\theta)\hat{\mathcal{P}}_z\hat{\mathcal{R}}_y(\theta)\hat{\mathcal{R}}_z(\varphi)$. Here, $\hat{\mathcal{R}}_a(\beta)$ is the rotation operator of the coordinate frame about the $a$-axis by the angle $\beta$. ($\hat{\mathcal{P}}_z = \mathrm{diag}(1,1,0)$ is a projector onto the *xy* plane). In the circular polarization basis, the transformation matrix can be formulated as (*1*):

$$\mathbf{E}(\mathbf{k}) \propto \begin{pmatrix} A & -Be^{-2i\varphi} & -Ce^{-i\varphi} \\ -Be^{2i\varphi} & A & -Ce^{i\varphi} \\ -Ce^{i\varphi} & -Ce^{-i\varphi} & 2B \end{pmatrix} \mathbf{E}_0 \quad (S3)$$

where $A = 1 + \cos^2\theta, B = \sin^2\theta$ and $C = \sqrt{2}\sin(2\theta)/2$ (Fig.S1D). The off-diagonal elements carry azimuthally dependent phase terms that exhibit *spin-to-orbit angular momentum conversion* associated with the geometric phases induced by the non-commutative 3D rotations of the global coordinate frame. This process can be formulated as $|\pm\sigma\rangle \rightarrow A|\pm\sigma\rangle - Be^{2i\sigma\varphi}|\mp\sigma\rangle$ (helicity $\sigma = \pm 1$) under illumination by circularly polarized waves(*2, 3*). By neglecting the longitudinal field component along to the *z*-axis, the upper left block matrix in Eq.(S3) acts as a dimensionally reduced operator $\widehat{\mathcal{H}}(\mathbf{k})$ for the in-plane field components, that is:

$$\widehat{\mathcal{H}}(\mathbf{k}) = \begin{pmatrix} A & -Be^{-2i\varphi} \\ -Be^{2i\varphi} & A \end{pmatrix} \quad (S4)$$

Now we consider a linear dipole $\mathbf{E}_0 = (e^{-i\Phi}, e^{i\Phi}, 0)^T = (e^{-i(\varphi+\xi)}, e^{i(\varphi+\xi)}, 0)^T$ located off-origin in the *xy* plane at $\mathbf{r} = r(\cos\varphi, \sin\varphi, 0)$, with a polarization angle $\xi \in [0, \pi)$ relative to the radial **r**-axis in Fig.S1A. The scattered field at an observation point $(0,0,f)$ can be expressed as:

$$\mathbf{E} \propto \begin{pmatrix} Ae^{-i\Phi} - Be^{-2i(\varphi+\pi)+i\Phi} \\ Ae^{+i\Phi} - Be^{2i(\varphi+\pi)-i\Phi} \\ -Ce^{i(\varphi+\pi)}e^{-i\Phi} - Ce^{-i(\varphi+\pi)}e^{i\Phi} \end{pmatrix} = \begin{pmatrix} \mathcal{S}e^{-i\varphi} \\ \mathcal{S}^*e^{i\varphi} \\ 2C\cos\xi \end{pmatrix} \quad (S5)$$

Here, $\mathcal{S}(\xi) = Ae^{-i\xi} - Be^{+i\xi}$ serves as the scattering amplitude for the generated in-plane components, quantifying both the relative polarization orientation of sources and the strength of SOI. It is notable that $\mathcal{S}$ exhibits Hermitian conjugation with respect to $\xi$, i.e., $\mathcal{S}(\xi) = \mathcal{S}^*(-\xi)$. When $\theta \rightarrow 0$, $|\mathcal{S}| = $ const and $\arg\{\mathcal{S}\} \propto \xi$, indicating that the polarization of propagation wave is fully determined by the source, namely, symmetry inheritance. As $\theta$ increases, the profiles of $|\mathcal{S}|$ and $\arg\{\mathcal{S}\}$ are progressively distorted by SOI. At $\theta$ approaching $\pi/2$ (grazing incidence limit), $|\mathcal{S}| \propto \sin\xi$ and $\arg\{\mathcal{S}\} = \pi/2$ (Fig.S1B-C). This implies that the scattered field's polarization becomes source-independent ($\xi$ only modulates the amplitude) and consequently exhibits robustness in the strong SOI regime. In contrast, the generated longitudinal component $E_z \propto C\cos(\xi)$ remains independent of the source orientation ("$e^{\pm i\varphi}$") during propagation (Fig.S1E).

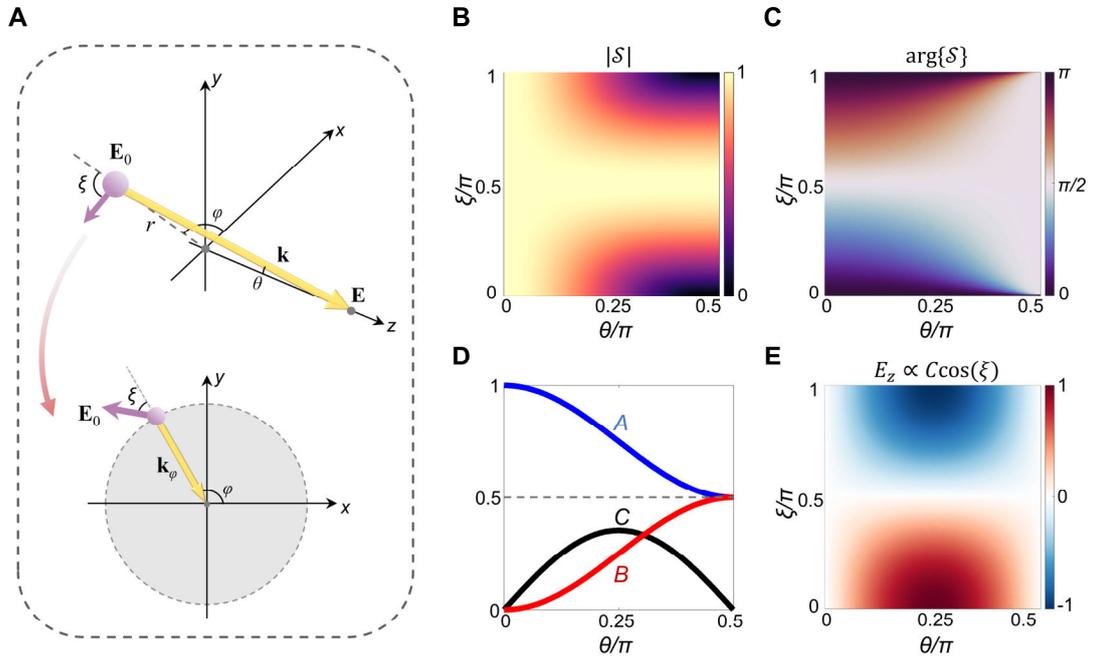

**Fig. S1. Scattered field of a linearly polarized dipole in the circular polarization basis.** (**A**) Schematic of a linear dipole positioned off the origin with in-plane polarization $\xi$ (purple arrows) and azimuth-dependent momentum $\mathbf{k}_\varphi$ (yellow arrows). (**B**) Magnitude $|\mathcal{S}|$ and (**C**) phase $\arg\{\mathcal{S}\}$ as functions of $\xi$ and $\theta$. (**D**) Coefficients of $A, B$ and $C$ versus $\theta$. (**E**) Dependence of the generated longitudinal field $E_z$ on $(\xi, \theta)$. Here the deflection angle $\theta = \operatorname{atan}(r/f)$.

# Section S2: Optical lattices formed by interference of $C_N$ dipoles and corresponding symmetry analysis

In this section, we derive the interference field of $C_N$ ($N \geq 3$) dipoles in the circular polarization basis and construct the corresponding Stokes parameters. We further analyze the symmetry properties of the generated optical lattices for both odd and even values of $N$.

The configuration of $C_N$ dipoles interference system is shown in Fig.S2A and 2B. Here, the linearly monochromatic dipoles $\mathbf{E}_i$ are located in the z = 0 plane at $\mathbf{r}_i = r(\cos\varphi_i, \sin\varphi_i, 0)$ with the azimuthal angles $\varphi_i = 2\pi(i-1)/N$, $i = \{1, \ldots, N\}$, and exhibit in-plane polarization angle $\Phi_i = \varphi_i + \xi_i$, with $\xi_i \in [0, \pi)$ defined as the angle relative to the radial vectors $\mathbf{r}_i$. This system possesses $C_N$ symmetry about the z-axis, and we consider the interference field $\mathbf{E}(x,y)$ at the plane $z = f$. If the observation region is significantly smaller than the radial distance $r$, the in-plane radiation momentum from each source can be approximated as $\mathbf{k}_i(x, y) \approx \mathbf{k}_i$ on the observation plane, where $\mathbf{k}_i = -k_0 \sin\theta(\cos\varphi_i, \sin\varphi_i) \propto \mathbf{r}_i$. Thus, the interference field at the observation plane can be formulated as:

$$\mathbf{E}(x,y) = \sum_{i=1}^{N} \widehat{\mathcal{H}}(\mathbf{k}_i)\mathbf{E}_i = \sum_{i=1}^{N} \begin{pmatrix} \mathcal{S}(\xi_i)e^{-i\varphi_i} \\ \mathcal{S}^*(\xi_i)e^{i\varphi_i} \end{pmatrix} e^{-ik_0\sin\theta(\cos\varphi_i x + \sin\varphi_i y)} \quad (S6)$$

We calculated the spatial distributions of various optical fields (e.g., Intensity and $S_3$) on the observation area. Results based on Eq.(S6) (Fig.S2D) show excellent agreement with dyadic Green's function (4, 5) (Fig.S2C), confirming the validity of the approximation $\mathbf{k}_i(x, y) \approx \mathbf{k}_i$. Furthermore, the Stokes parameters under the circular polarization basis can be obtained as:

$$\begin{cases} S_0 = |E_R|^2 + |E_L|^2 \\ S_1 = 2\Re\{E_R E_L^*\} \\ S_2 = -2\Im\{E_R E_L^*\} \\ S_3 = |E_R|^2 - |E_L|^2 \end{cases} \quad (S7)$$

where $E_R/E_L$ is the right- and left-handed circular polarization (RCP and LCP) components of the electric fields.

Substituting Eq.(S6) into Eq.(S7), then we obtain:

$$I_R(\mathbf{r}) = |E_R|^2 = \sum_{i=1}^{N}|S_i|^2 + \sum_{i\neq j}^{N}\Re\{S_iS_j^*\}\cos(\varphi_{ji} + \mathbf{k}_{ji}\cdot\mathbf{r}) \quad (S8)$$
$$-\sum_{i\neq j}^{N}\Im\{S_iS_j^*\}\sin(\varphi_{ji} + \mathbf{k}_{ji}\cdot\mathbf{r})$$

$$I_L(\mathbf{r}) = |E_L|^2 = \sum_{i=1}^{N}|S_i|^2 + \sum_{i\neq j}^{N}\Re\{S_iS_j^*\}\cos(\varphi_{ij} + \mathbf{k}_{ji}\cdot\mathbf{r}) \quad (S9)$$
$$+\sum_{i\neq j}^{N}\Im\{S_iS_j^*\}\sin(\varphi_{ij} + \mathbf{k}_{ji}\cdot\mathbf{r})$$

$$S_0(\mathbf{r}) = 2\sum_{i=1}^{N}|S_i|^2 + 2\sum_{i\neq j}^{N}\{\Re\{S_iS_j^*\}\cos(\varphi_{ij}) + \Im\{S_iS_j^*\}\sin(\varphi_{ij})\}\cos(\mathbf{k}_{ji}\cdot\mathbf{r}) \quad (S10)$$

$$S_1(\mathbf{r}) = 2\sum_{i,j}^{N}\{\Re\{S_iS_j\}\cos(\varphi_i + \varphi_j) + \Im\{S_iS_j\}\sin(\varphi_i + \varphi_j)\}\cos(\mathbf{k}_{ji}\cdot\mathbf{r}) \quad (S11)$$
$$+2\sum_{i,j}^{N}\{\Re\{S_iS_j\}\sin(\varphi_i + \varphi_j) - \Im\{S_iS_j\}\cos(\varphi_i + \varphi_j)\}\sin(\mathbf{k}_{ji}\cdot\mathbf{r})$$

$$S_2(\mathbf{r}) = 2\sum_{i,j}^{N}\{\Re\{S_iS_j\}\sin(\varphi_i + \varphi_j) - \Im\{S_iS_j\}\cos(\varphi_i + \varphi_j)\}\cos(\mathbf{k}_{ji}\cdot\mathbf{r}) \quad (S12)$$
$$-2\sum_{i,j}^{N}\{\Re\{S_iS_j\}\cos(\varphi_i + \varphi_j) + \Im\{S_iS_j\}\sin(\varphi_i + \varphi_j)\}\sin(\mathbf{k}_{ji}\cdot\mathbf{r})$$

$$S_3(\mathbf{r}) = 2\sum_{i\neq j}^{N}\{\Re\{S_iS_j^*\}\sin(\varphi_{ij}) - \Im\{S_iS_j^*\}\cos(\varphi_{ij})\}\sin(\mathbf{k}_{ji}\cdot\mathbf{r}) \quad (S13)$$

Here, $\varphi_{ji} = 2\pi(j - i)/N$. Notably, the set of radiation momentum differences $\mathbf{k}_{ji} = \mathbf{k}_j - \mathbf{k}_i, i, j = \{1, \ldots, N\}, i \neq j$ possesses $N$-fold rotational symmetry (i.e. $\hat{\mathcal{R}}_{2\pi/N}\mathbf{k}_i = \mathbf{k}_{i+1}, i = \{1, \ldots, N\}$), serving as reciprocal lattice vectors in the interference system. In two dimensions, periodic Bravais lattices are permitted exclusively when $N = 3,4,6$, corresponding to triangular/hexagonal ($N = 3$ or $6$) and square ($N = 4$) lattices (6), whereas other cases (e.g., $N = 5,7,8,9\ldots$) or even higher rotational symmetries are only allowed to generate quasicrystals textures due to the incommensurability of $\{\mathbf{k}_{ji}\}$ sets as shown in Fig.S3 and S4. We further construct $C_N$ dipole sources by enforcing $\xi_i = \xi$. The expression of corresponding optical fields simplifies to:

$$\begin{cases} I_R(\mathbf{r}) \propto \sum_{i \neq j}^{N} \cos(\varphi_{ji} + \mathbf{k}_{ji} \cdot \mathbf{r}) \\ I_L(\mathbf{r}) \propto \sum_{i \neq j}^{N} \cos(\varphi_{ij} + \mathbf{k}_{ji} \cdot \mathbf{r}) \\ I_T(\mathbf{r}) \propto \sum_{i \neq j}^{N} \cos(\varphi_{ji}) \cos(\mathbf{k}_{ji} \cdot \mathbf{r}) \\ S_3(\mathbf{r}) \propto \sum_{i \neq j}^{N} \sin(\varphi_{ji}) \sin(\mathbf{k}_{ji} \cdot \mathbf{r}) \end{cases} \quad (S14)$$

It is naturally that the spatial profiles of the above optical fields also exhibits $N$-fold symmetry (e.g., $\hat{\mathcal{R}}_{2\pi/N} I_R(\mathbf{r}) = I_R(\hat{\mathcal{R}}_{2\pi/N}^{-1}\mathbf{r}) = I_R(\mathbf{r})$). Moreover, the total intensity and the longitudinal spin component exhibit even and odd parity, respectively, meaning:

$$\begin{cases} \hat{P} I_T(\mathbf{r}) = I_T(-\mathbf{r}) = I_T(\mathbf{r}) \\ \hat{P} S_3(\mathbf{r}) = S_3(-\mathbf{r}) = -S_3(\mathbf{r}) \end{cases} \quad (S15)$$

Combined with the $N$-fold symmetry, $S_3(\mathbf{r}) \equiv 0$ when $N$ is even. This implies that the constructed Stokes vectors field can only form in-plane vortex structures (Fifth panel in Fig.S4) rather than 3D topological textures. Therefore, we focus on the cases with odd values of $N$. In particular, $N = 3$ is the unique case that forms non-zero periodic Stokes textures, with the lattice constant $a = 4\pi/(\sqrt{3}|\mathbf{k}_{ji}|) = 2\lambda/(3\sin\theta), i \neq j$.

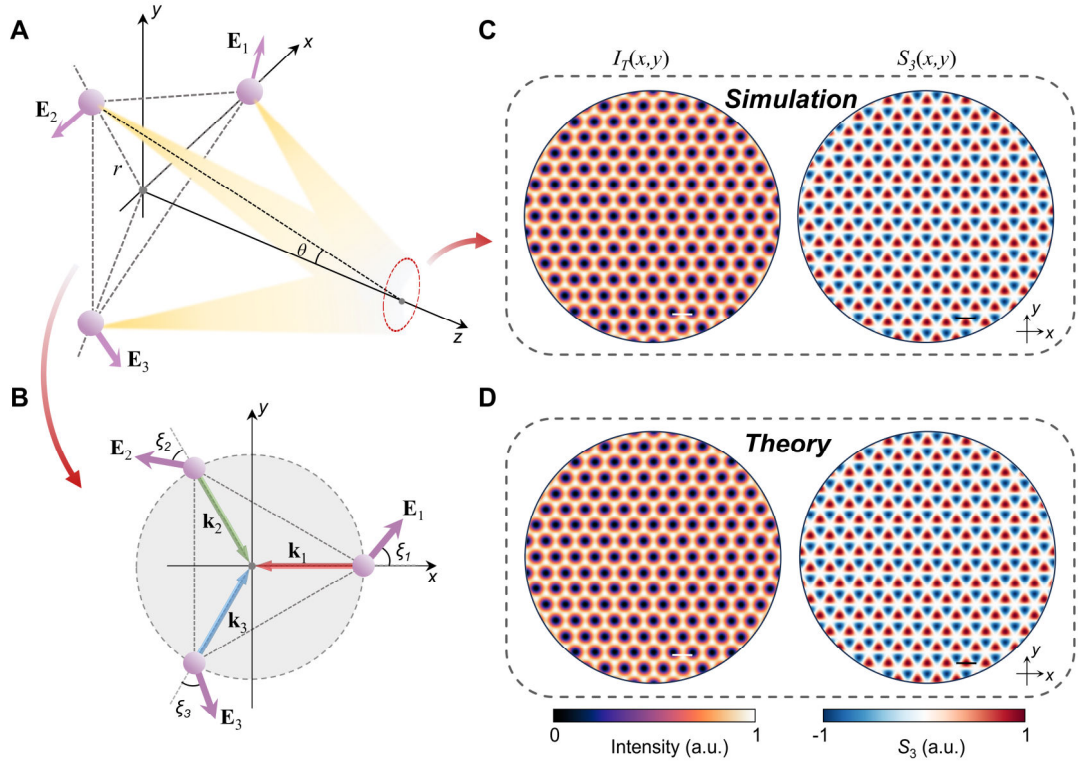

**Fig. S2. Schematic of the multi-dipole interference system.** (**A**) 3D view of the interference process. (**B**) Spatial arrangement and polarization configurations of dipoles in the $z = 0$ plane. Interference intensity and spin distribution on the observation plane $z = f$, obtained from (**C**) rigorous calculations and (**D**) theory based on Eq.(S6). respectively. The scale bar is the lattice constant $a = 2\lambda/(3\sin\theta)$ ($\lambda$ is the free-space wavelength). $N$ is set to 3 and $\theta = 0.1\pi$. Distance $r = 100\lambda$.

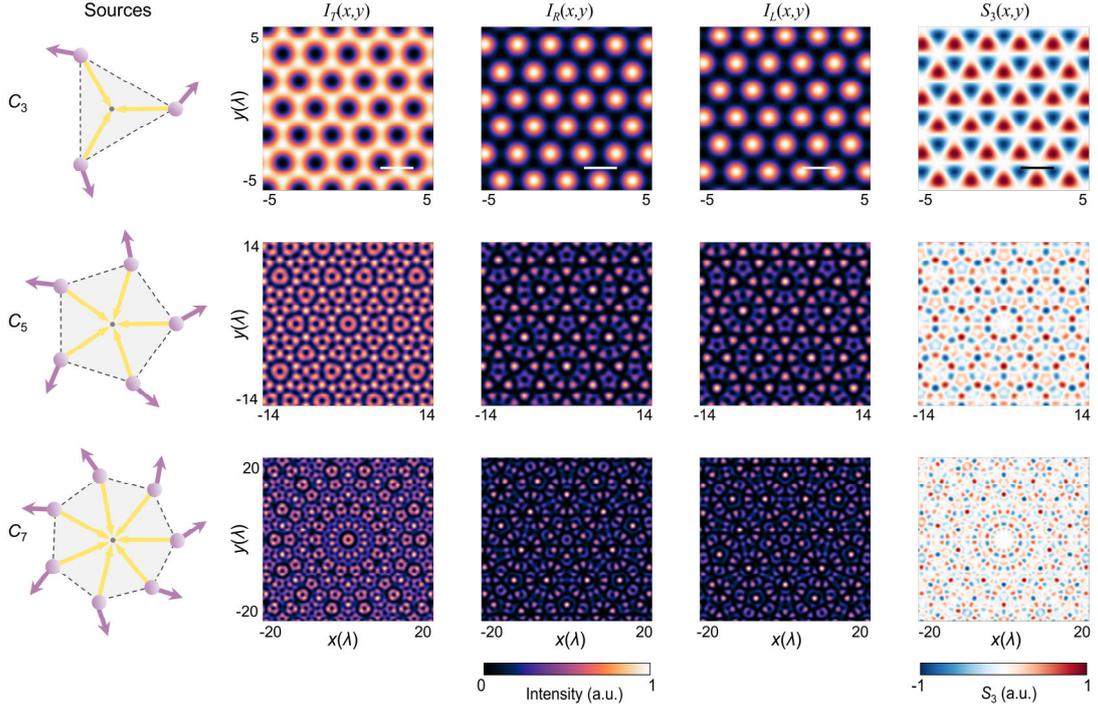

**Fig. S3. Optical lattices generated by $C_N$ dipole sources with odd $N = 3,5,7$.** Panels from left to right: multi-dipole with $N$-fold rotational in-plane momenta (yellow arrows) and polarization vectors (purple arrows), the total intensity $I_T$, the intensities of RCP/LCP components $I_R$ and $I_L$, and the Stokes $S_3$.

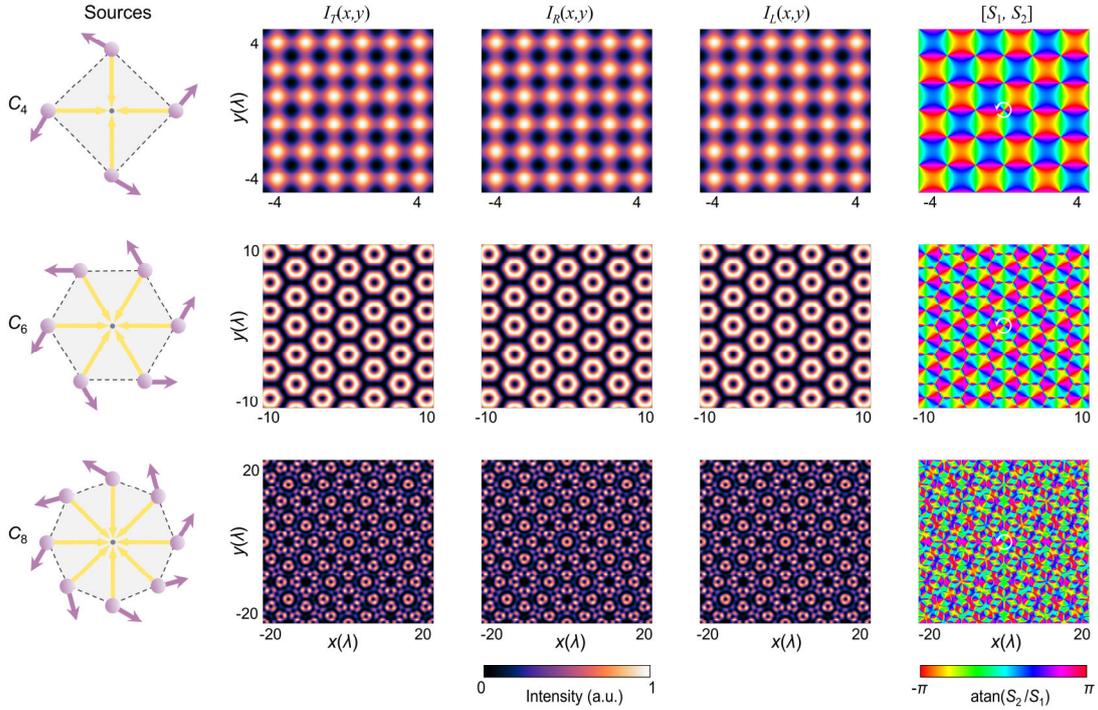

**Fig. S4. Cases for even $N = 4,6,8$.** The last column of panels shows the phase distribution of non-zero Stokes $S_1$ and $S_2$.

## Section S3: Topological properties of Stokes $C_3$ merons

In the section, we performed a detailed calculation of the topological properties of ideal Stokes $C_3$ meron lattices corresponding to the basic solutions in Fig.2A, including the skyrmion number ($N_{sk}$), vorticity ($m$) and helicity ($\gamma$) with varying input polarization configurations, etc. A general normalized Stokes vector field $\mathbf{s}(x,y) = [s_1, s_2, s_3]$ in the Cartesian coordinate can be mapped onto the Poincaré sphere, with the skyrmion number $N_{sk} = (1/4\pi)\iint_A \rho(x,y)dxdy$ ($A$ denotes the region of a unitcell of the lattice) and $\rho(x,y)$ represents the skyrmion density, that is:

$$\rho(x,y) = \mathbf{s}(x,y) \cdot [\partial_x \mathbf{s}(x,y) \times \partial_y \mathbf{s}(x,y)] \tag{S16}$$

Here the sign of $\rho(x,y)$ denotes the sense of the vector field wrapping the unit sphere. We calculated the distribution of $\rho(x,y)$ for a perfect $C_3$ meron lattice as shown in Fig.S5A, exhibiting the alternating equilateral triangular distributions similar to $s_3(x,y)$. Notably, the skyrmion density gradually diverges at the edges of triangular cells as the light intensity approaches zero. Since the skyrmion number cannot uniquely determine the nontrivial vector textures, we map $\mathbf{s}(x,y) = \mathbf{s}(r\cos\phi, r\sin\phi) = (\sin\beta\cos\alpha, \sin\beta\sin\alpha, \cos\beta)$ onto a unit sphere, where $r$ and $\phi$ are the radial and azimuthal coordinates of the vector field in $xy$ plane. Thus, $N_{sk}$ can be written as:

$$\begin{aligned} N_{sk} &= \frac{1}{4\pi} \int_0^{r_0} dr \int_0^{2\pi} d\phi \frac{d\beta(r)}{dr} \frac{d\alpha(\phi)}{d\phi} \sin\beta(r) \\ &= \frac{1}{4\pi} [\cos\beta(r)]_{r=0}^{r=r_0} [\alpha(\phi)]_{\phi=0}^{\phi=2\pi} \end{aligned} \tag{S17}$$

The number of vorticity $m$ and helicity $\gamma$ can be defined as (7):

$$m = \frac{1}{2\pi} \oint_C d\phi = \frac{1}{2\pi} [\alpha(\phi)]_{\phi=0}^{\phi=2\pi} \tag{S18}$$

$$\gamma = \alpha(\phi) - m\phi \tag{S19}$$

Based on above indices, the 2D topological quasiparticles can be classified into several basic configurations as shown in Fig.S6. For vorticity $m = 1$, the cases of $\gamma = 0$ and $\pi$ are Néel type textures characterized by a hedgehog-like distribution, and $\gamma = \pm\pi/2$ are Bloch types with in-plane vortex distribution (Top panels in Fig.S6). In the case of $m = -1$, such configurations are defined as the anti-type structures featuring a saddle texture (anti-vortex), which reside in a different topologically protected group (8) form both Bloch and Néel type skyrmionics textures. (Bottom panels in Fig.S6).

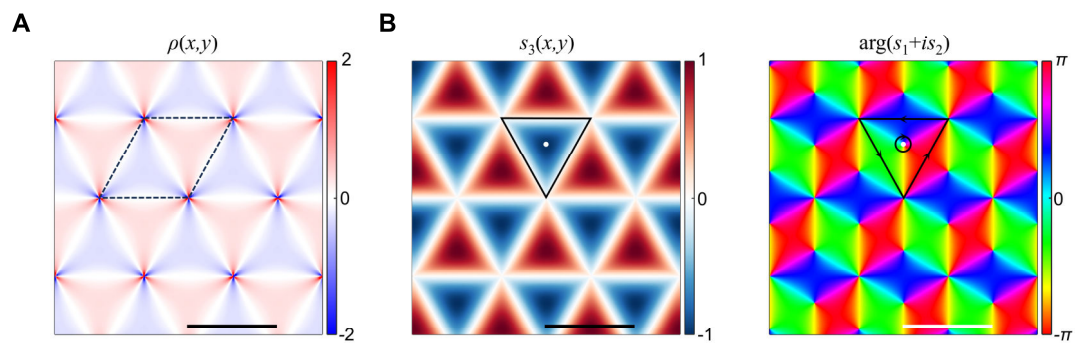

**Fig. S5. Details about calculation of Skyrmion number for ideal $C_3$ merons.** (**A**) Skyrmion density distributions for a ideal $C_3$ meron lattice. The black dashed line marks a unit cell of the texture that enclose a meron pair. (**B**) Distribution of normalized Stokes $s_3$ and the $\arg(s_1 + is_2)$. Two black loops with opposite handedness enclose the singularity (white dot) and the boundary of a single meron.

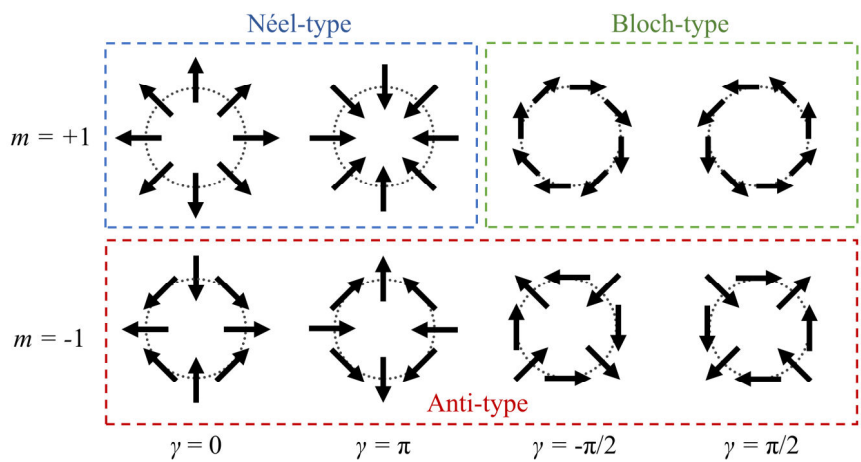

**Fig. S6. Typical transverse vector configurations of 2D topological quasiparticles with various values of $m$ and $\gamma$.**

To precisely capture the skyrmion number $N_{sk}$ within a finite region, we employ a geometric approach proposed in Ref.(9) instead of the surface integral method, that is:

$$N_{sk} = \sum_j \frac{1}{4\pi} \oint_{\beta_j} s_z \nabla \Phi \cdot d\boldsymbol{l} - \frac{1}{4\pi} \oint_\alpha s_z \nabla \Phi \cdot d\boldsymbol{l} \tag{S20}$$

where $s_z = s_3$ and $\Phi = \arg(s_1 + is_2)$. Equation S20 reformulates the skyrmion number as a sum of two closed line integrals: a counter-clockwise ($\alpha$) loop along the boundary of selected region, and a clockwise loop ($\beta_j$) that encloses around each singularity of $\Phi$, as shown in Fig.S5B. The numerical error of $N_{sk}$ calculated with Eq.(S20) is about 1%.

We further systematically calculated $N_{sk}, m$, and $\gamma$ for the four basic solutions shown in Fig.2A in the absence of SOI. Detailed information of the four types of solutions and their associated symmetry groups are summarized in Table S1. As shown in Fig.S7A, the skyrmion number $N_{sk} \approx \pm 0.49$ remains stable, demonstrating the topological robustness of Stokes merons formed through dipole interference. Moreover, the solutions of Type I, which belong to $C_3$ group, exclusively yield anti-type merons ($m = -1$). On the other hand, the configurations of Type II-IV (anti-$C_3$ group) produce merons with positive vorticity ($m = 1$), whose helicity is continuously modulated by initial polarization states, i.e., $\gamma \propto t$, leading to an evolution of merons from Bloch, Intermediate (twisted textures), to Néel types (Fig.S7B-E).

**Table S1. Summary of polarization groups corresponding to exact symmetry operations in the absence of SOI.**

| Type | Parametric equations of polarization states | Symmetry group |
|---|---|---|
| I | $t \in [0, \pi); \xi_1 = t; \xi_2 = t; \xi_3 = t;$ | $C_3$ |
| II | $t \in [0, \pi/3]; \xi_1 = t + \pi/3; \xi_2 = t; \xi_3 = t + 2\pi/3;$ | Anti-$C_3$ |
| III | $t \in [0, \pi/3]; \xi_1 = t + 2\pi/3; \xi_2 = t + \pi/3; \xi_3 = t;$ | Anti-$C_3$ |
| IV | $t \in [0, \pi/3]; \xi_1 = t; \xi_2 = t + 2\pi/3; \xi_3 = t + \pi/3;$ | Anti-$C_3$ |

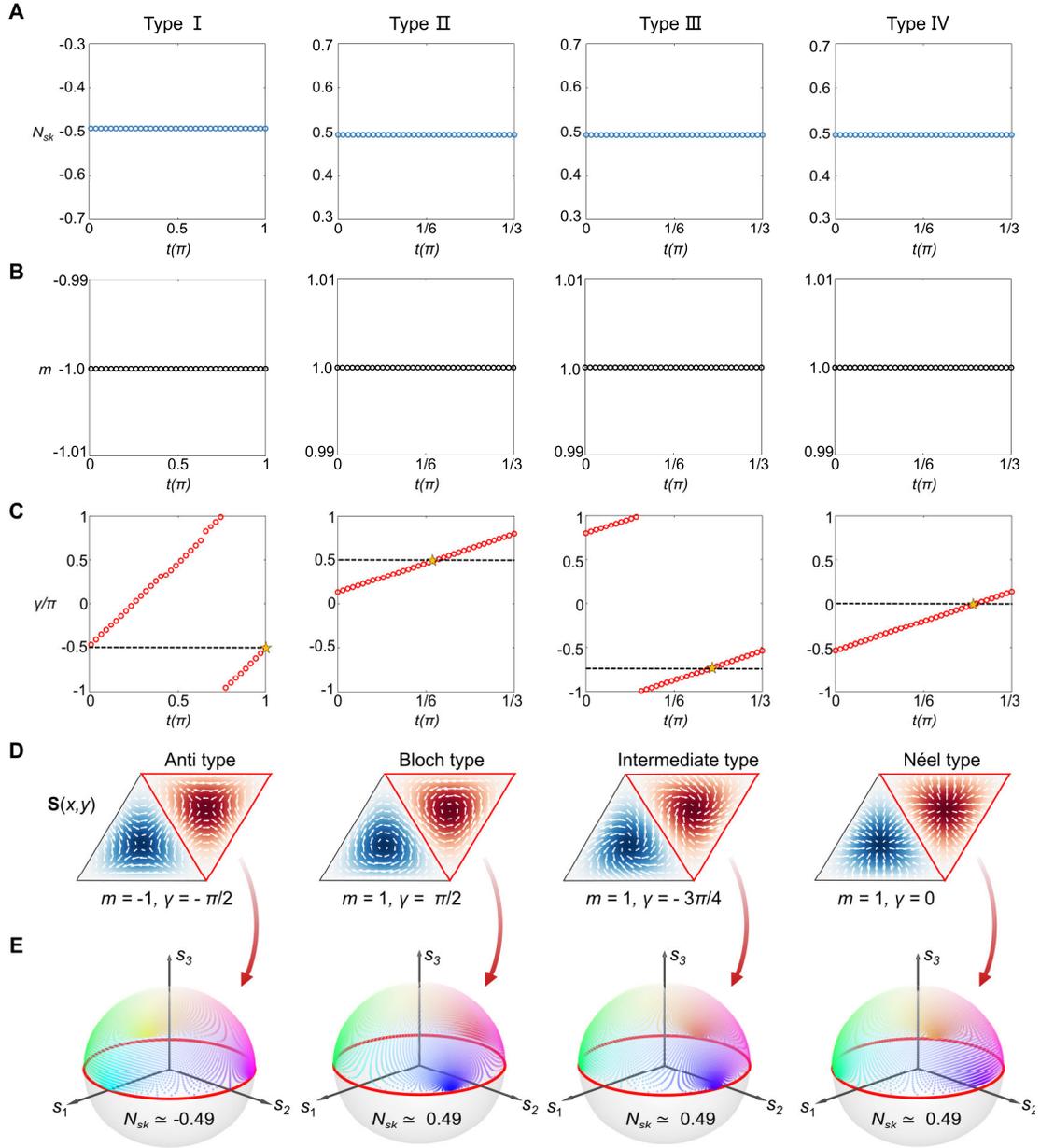

**Fig. S7. Topological features for ideal merons corresponding to the polarization configurations in the absence of SOI.** (**A**) Calculated skyrmion number $N_{sk}$, (**B**) vorticity $m$, (**C**) helicity $\gamma$ as functions of input polarization states $t$, and (**D**) selected Stokes vectors $\mathbf{S}(x,y)$ over a triangular cell (yellow stars in (C)) together with (**E**) the mapping onto the Poincaré spheres.

## Section S4: Derivation of the overlapping factor $\mathcal{K}$

In this section, we define the overlapping factor $\mathcal{K}$ as a metric to evaluate the *N*-fold symmetry of optical lattices in real space; we derive its explicit expression based on the formalism of Eq.(S6)—(S14). Furthermore, we calculate the normalized density of states in polarization space $(\xi_1, \xi_2, \xi_3)$ for ideal $C_3$ lattices spanning different SOIs.

To quantify the degree of lattice distortion, we utilize the normalized Fourier spectrum to characterize the translational and rotational symmetry of interference lattices, thereby eliminating the ambiguities caused by lattice's misalignment in the real space. As shown in Fig.S8, the Fourier transform of the three-wave interference field that generates the perfect optical lattice (Fig.S8A) exhibits six peaks of identical intensity in momentum space (Fig.S8C), while the Fourier peaks of a distorted lattice (Fig.S8B) may be attenuated or even vanished (Fig.S8D). Indeed, the magnitudes of six Fourier peaks in momentum space correspond to the complex amplitudes along the three propagation directions $\mathbf{k}_{ji} = \mathbf{k}_j - \mathbf{k}_i, \{i \neq j, i,j = 1,2,3\}$. Therefore, we introduce a normalized overlapping factor, that is:

$$\mathcal{K} = \frac{\iint |\mathcal{F}(S_3)| |\mathcal{F}(S_3^{\mathrm{p}})| \, dk_x dk_y}{\sqrt{\iint |\mathcal{F}(S_3)|^2 dk_x dk_y} \sqrt{\iint |\mathcal{F}(S_3^{\mathrm{p}})|^2 dk_x dk_y}} \quad (S21)$$

Here, $S_3$ denotes the distribution of a particular spin lattice, while $S_3^{\mathrm{p}}$ corresponds to the perfect triangular lattice defined by Eq.(S14). When $\mathcal{K}$ approaches unity, it indicates the emergence of an ideal $C_3$ lattice.

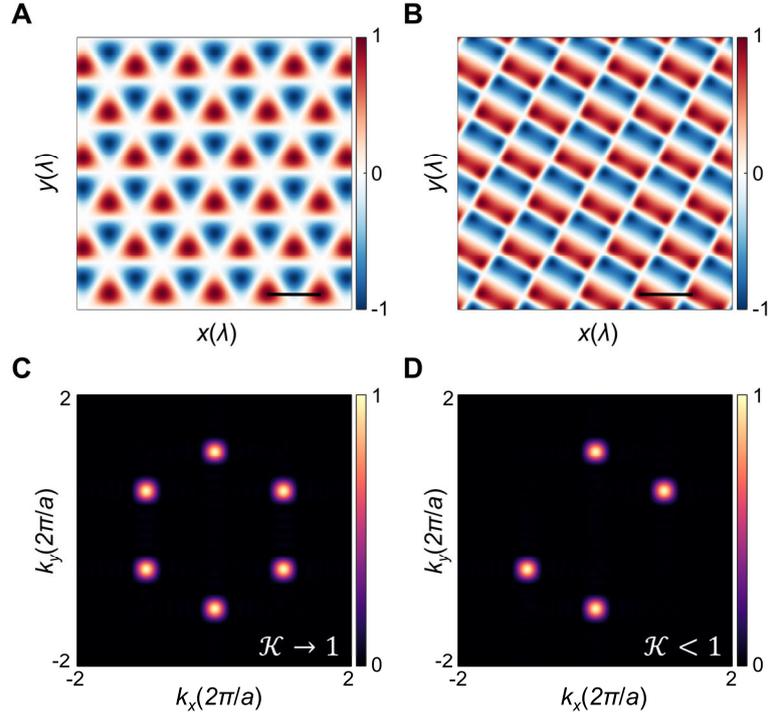

**Fig. S8. Comparison between optical lattices and their Fourier spectrum for both perfect and broken $C_3$ symmetry.** **(A)-(B)** Spin lattices with exact and broken $C_3$ symmetry in a finite region. **(C)-(D)** Corresponding Fourier transform results of $S_3$.

We further derive the formulation of $\mathcal{K}$ as a function of polarization combinations $(\xi_1, \xi_2, \xi_3)$. Considering the Fourier transform of a finite square lattice:

$$|\mathcal{F}(S_3)| \propto \left| \sum_{i \neq j}^{3} \{\Re\{S_i S_j^*\} \sin(\varphi_{ij}) - \Im\{S_i S_j^*\} \cos(\varphi_{ij})\} \mathcal{F}\{\sin(\mathbf{k}_{ji} \cdot \mathbf{r}) \, rect_l(\mathbf{r})\} \right|$$

$$= \left| \sum_{i \neq j}^{3} \{\Re\{S_i S_j^*\} \sin(\varphi_{ij}) - \Im\{S_i S_j^*\} \cos(\varphi_{ij})\} \mathcal{F}\{\sin(\mathbf{k}_{ji} \cdot \mathbf{r})\} \otimes \mathcal{F}\{rect_l(\mathbf{r})\} \right|$$

$$= \left| \sum_{i \neq j}^{3} \{\Re\{S_i S_j^*\} \sin(\varphi_{ij}) - \Im\{S_i S_j^*\} \cos(\varphi_{ij})\} [g(\mathbf{k} - \mathbf{k}_{ji}) - g(\mathbf{k} + \mathbf{k}_{ji})] \right|$$

Neglecting spatial overlap between Fourier peaks ($l \gg a$):

$$\approx \left| \sum_{i \neq j}^{3} \{\Re\{\mathcal{S}_i \mathcal{S}_j^*\} \sin(\varphi_{ij}) - \Im\{\mathcal{S}_i \mathcal{S}_j^*\} \cos(\varphi_{ij})\} \right| |g(\mathbf{k} \pm \mathbf{k}_{ji})| \quad \text{(S22)}$$

$$\propto \sum_{i \neq j}^{3} |\mathcal{S}_i \mathcal{S}_j^*| |\sin(\varphi_{ji} + \beta_{ij})| |g(\mathbf{k} \pm \mathbf{k}_{ji})|$$

Here, $g(\mathbf{k}) = sinc_l(\mathbf{k}) = sinc(lk_x/2) sinc_l(lk_y/2)$, $l$ is the side length of the selected region and $\beta_{ij} = \arg(\mathcal{S}_i \mathcal{S}_j^*)$. Similarly, for perfect lattices generated by $C_3$ sources:

$$|\mathcal{F}(S_3^p)| \propto \sum_{i \neq j}^{3} |\sin(\varphi_{ji})| |g(\mathbf{k} \pm \mathbf{k}_{ji})| \quad \text{(S23)}$$

Consider the integral relation:

$$\iint |g(\mathbf{k} \pm \mathbf{k}_{ji})| |g(\mathbf{k} \pm \mathbf{k}_{mn})| dk_x dk_y \propto \begin{cases} 0, ji \neq mn \\ \\ 1/l^2, ji = mn \end{cases} \quad \text{(S24)}$$

Substitute Eq.(S22)-(S24) into Eq.(S21), we finally obtain:

$$\mathcal{K}(\xi_1, \xi_2, \xi_3) = \frac{\sum_{i \neq j}^{3} |\mathcal{S}_i \mathcal{S}_j^*| |\sin(\varphi_{ji} + \beta_{ij})|}{\sqrt{3} \sqrt{\sum_{i \neq j}^{3} |\mathcal{S}_i \mathcal{S}_j^*|^2 |\sin(\varphi_{ji} + \beta_{ij})|^2}} \quad \text{(S25)}$$

Based on Eq.(S25), we calculated the density of states $\rho(\mathcal{K} > K_c)$, which stands for the percentage of polarization states in the polarization cube to produce perfect $C_3$ lattices. In Fig.S9A, $\rho(\mathcal{K} > K_c)$ rises steeply when $\theta$ exceeds $0.25\pi$ and reaches its maximum as $\theta \to \pi/2$, with the threshold $K_c$ set to 0.99. This phenomenon originates from the proliferation of q-$C_3$ groups under strong SOI. It should be noted that here we evaluate only the $\mathcal{K}$ factor of Stokes $S_3$ as a metric. Nevertheless, $\mathcal{K}(S_3) \to 1$ does not always guarantee the formation of $C_3$ merons (may be a coincidence caused by intensity distortion). To ensure the generation of ideal merons, $\mathcal{K}(S_0) \to 1$ must also be satisfied, which causes a slight difference in the evolution of polarization states. More details are provided in the movies.

For a q-$C_3$ states featuring the polarization configurations $(-\xi, \xi, \xi)$, the Hermitian conjugation that yields $|\mathcal{S}(\xi)\mathcal{S}^*(-\xi)| = |\mathcal{S}(\xi)| \cdot |\mathcal{S}^*(-\xi)| = |\mathcal{S}(\xi)|^2$, and Eq.(S25) can be further simplified as:

$$\mathcal{K}(-\xi, \xi, \xi; \theta) = \frac{\sum_{i \neq j}^{3} |\sin(\varphi_{ji} + \beta_{ij})|}{\sqrt{3}\sqrt{\sum_{i \neq j}^{3} |\sin(\varphi_{ji} + \beta_{ij})|^2}} \quad (S26)$$

As shown in Fig.S9B, $\mathcal{K}$ rapidly tends to unity across the entire continuous polarization ensemble as $\theta \to \pi/2$, and the proliferation effect remains stable for different values of $K_c$.

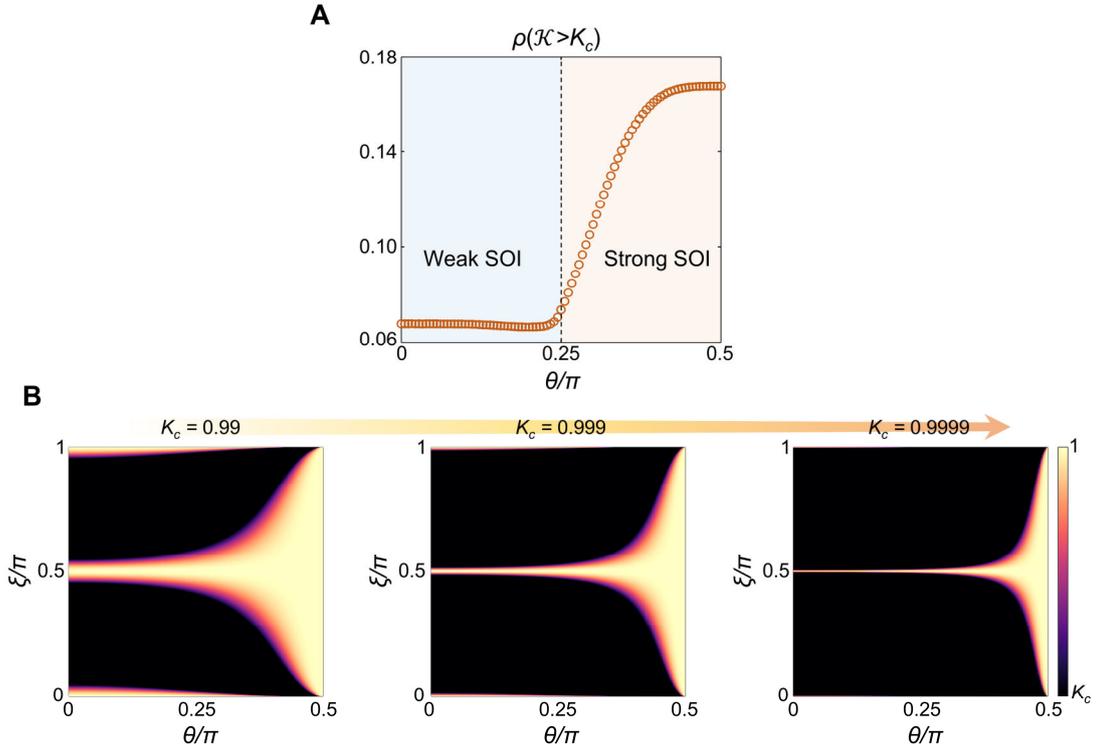

**Fig. S9.** (**A**) Calculated total density of polarization states for ideal $C_3$ lattices and (**B**) $\mathcal{K}$ profiles under different thresholds $K_c$ for the q-$C_3$ group.

## Section S5: Extension to the arbitrary polarization dipole sources

In this section, we extend the concept of quasi-symmetry from linearly polarized to circularly polarized, and arbitrary elliptically polarized dipoles interference systems. For the general in-plane polarization states $\mathbf{E}_i$ characterized by the azimuth angles $\Phi_i = \varphi_i + \xi_i$ and ellipticity $\chi_i \in [-\pi/4, \pi/4]$, the corresponding electric fields can be expressed as:

$$\mathbf{E}_i = \frac{1}{\sqrt{2}}\begin{pmatrix} 1 & -i \\ 1 & i \end{pmatrix}\begin{pmatrix} \cos\Phi_i & -\sin\Phi_i \\ \sin\Phi_i & \cos\Phi_i \end{pmatrix}\begin{pmatrix} \cos\chi_i \\ i\sin\chi_i \end{pmatrix} = \frac{1}{\sqrt{2}}\begin{pmatrix} e^{-i\Phi_i}(\cos\chi_i + \sin\chi_i) \\ e^{+i\Phi_i}(\cos\chi_i - \sin\chi_i) \end{pmatrix} \quad (S27)$$

Similarly, the corresponding scattered field is:

$$\mathbf{E}(\mathbf{k}) \propto \widehat{\mathcal{H}}(\mathbf{k})\mathbf{E}_i = \begin{pmatrix} S_R e^{-i\varphi_i} \\ S_L e^{+i\varphi_i} \end{pmatrix}$$

where

$$\begin{aligned} S_R &= Ae^{-i\xi_i}(\cos\chi_i + \sin\chi_i) - Be^{+i\xi_i}(\cos\chi_i - \sin\chi_i) \\ S_L &= Ae^{+i\xi_i}(\cos\chi_i - \sin\chi_i) - Be^{-i\xi_i}(\cos\chi_i + \sin\chi_i) \end{aligned} \quad (S28)$$

It is noteworthy that the Hermitian conjugation is invalid between the RCP and LCP components of the scattering amplitudes (i.e., $S_R \neq S_L^*$), due to the presence of non-zero ellipticity $\chi$, leading to different results from the linearly polarization cases.

Firstly, we consider circularly polarized $C_3$ dipoles with the ellipticity $\chi = \pm\pi/4$ as shown in Fig.S10A and D. When $\theta \approx 0$, the trivial hexagonal lattices are formed (Fig.S10C and F's top panels), exhibiting the spatial distribution of the pure $S_3$ components (up or down) that maps onto the two poles of the Poincaré sphere. It means that the field inherits the spin states of sources via free-space transport. As $\theta$ increases, spin-flipped components arise from SOI, leading the Stokes vectors to progressively diffuse from the poles toward the equator. When $\theta \approx \pi/2$, ideal $C_3$ Stokes merons are generated, covering the entire area of hemisphere as a result of extreme spin conversion induced by strongest SOI (Bottom panels in Fig.S10C and F). Calculated skyrmion numbers ($N_{sk}$) evolve continuously from 0 to $\pm 0.5$ with $\theta$, confirming the SOI-driven topological transition as shown in Fig.S10B and E.

Then we focus on the case of elliptically polarized dipoles. For simplicity, we set the ellipticity of the three dipoles to be identical, i.e. $\chi_i = \chi \in [-\pi/4, \pi/4], i = 1,2,3$. (ensure the well-defined symmetry groups). We first calculate $N_{sk}$ as a function of ellipticity $\chi$ and $\theta$ through the Stokes lattices generated from the $C_3$ group (left panel, Fig.S11A). It can be observed that in the paraxial approximation ($\theta \to 0$), the formation of topological textures with $\chi \neq 0$ is forbidden. As $\theta$ increases, this restriction is gradually lifted. When $\theta \gtrsim 0.45\pi$, dipole configurations with arbitrary ellipticity are permitted to generate meron textures due to strong SOI. This striking result indicates that the intrinsic SOI not only suppresses the influence of relative polarization ($\xi$) but also eliminate the requirement on the ellipticity ($\chi$) of the light sources, thereby giving rise to a highly robust polarization zone for sources to induce topological lattices. Moreover, as SOI is enhanced, the influence of ellipticity $\chi$ on the density of polarized states for perfect lattices becomes more pronounced—i.e., larger $|\chi|$ corresponds to higher $\rho$ (Fig.S11B). The underlying mechanism is that larger ellipticity reduces the effect of polarization orientation. When $\theta \to \pi/2$, $\rho$ rapidly converges to unity under purely circular-polarized illumination ($\chi = \pm\pi/4$). At this stage, $\xi$ will introduce different initial phase delays, which only leads to the displacement rather than distortion of the lattices in the case of $N = 3$ (*10*).

Finally, we discuss the behaviors of the polarization symmetry groups. When $|\chi| > 0$, the symmetry-guaranteed property of $C_3$ group is progressively broken, which prohibits the formation of ideal meron textures under weak SOI even the sources exhibit global rotational symmetry. The q-$C_3$ group also undergoes a similar transition. As $|\chi|$ increases toward $\pi/4$, the $C_3$ and q-$C_3$ groups become indistinguishable due to the redundancy of the partial mirror operations under circular polarized states (Fig.S12). Indeed, strong SOI ensures that:

$$\left|S_R\left(\xi;\theta\to\frac{\pi}{2}\right)\right|=\left|S_L\left(\pm\xi;\theta\to\frac{\pi}{2}\right)\right| \qquad (S29)$$

This relation demonstrates that $C_3$ and q-$C_3$ serve as the fundamental configurations across an extensive parameter space, and further highlights the essential role of SOI.

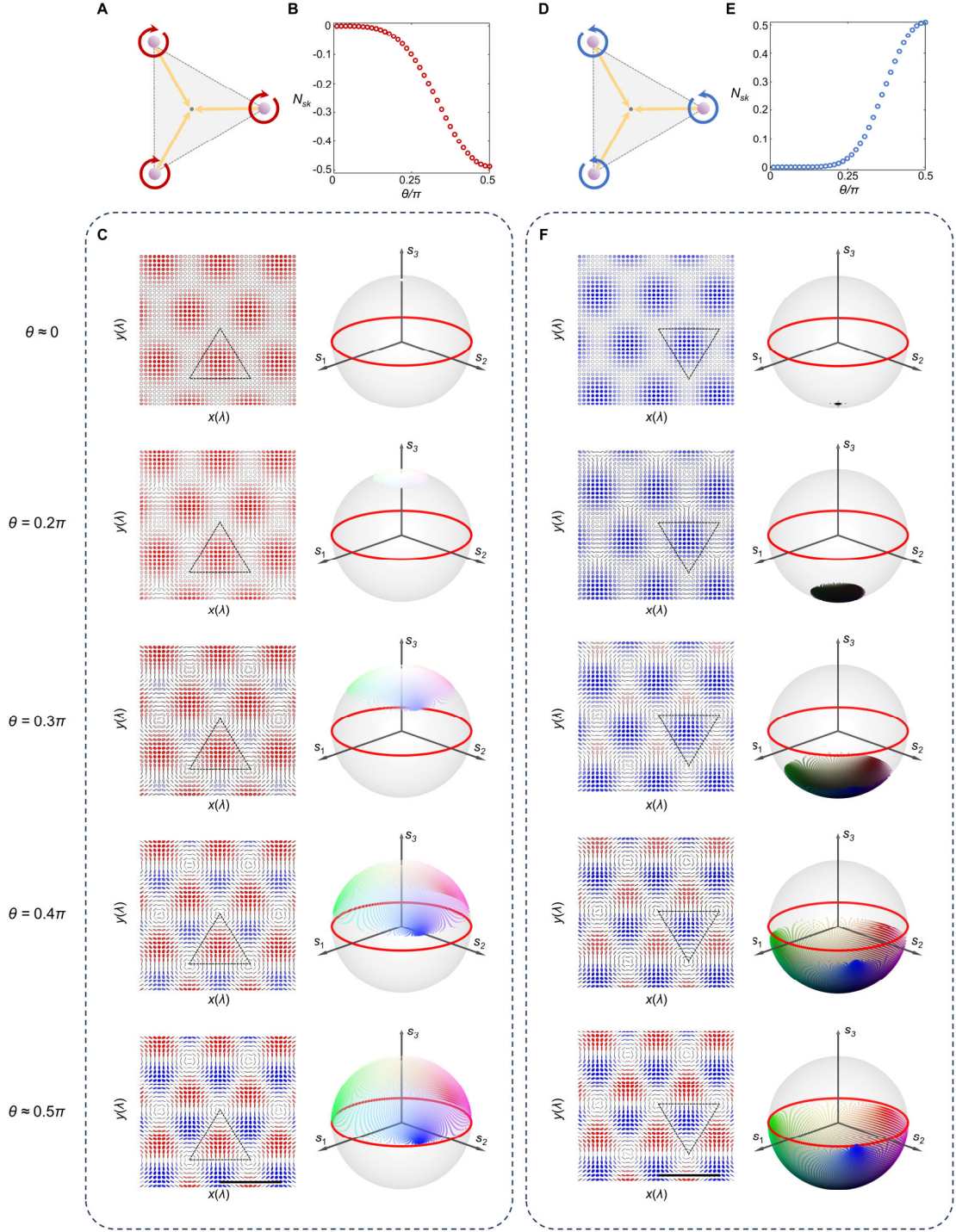

**Fig. S10. Topological transition from trivial to meron lattices induced by strong SOI under circular-polarized dipole sources.** Schematic of in-plane (**A**) RCP and (**D**) LCP dipoles located in the *xy* plane. Calculated skyrmion number $N_{sk}$ as functions of $\theta$ for (**B**) $\chi = \pi/4$ and (**E**) $\chi = -\pi/4$. (**C**)(**F**) Typical polarization ellipse distribution and corresponding the mapping of normalized Stokes vectors over identical triangular

unitcells (black dashed boundary) onto the unit sphere. In the lattices, the filled color of ellipses represents the values of normalized spin.

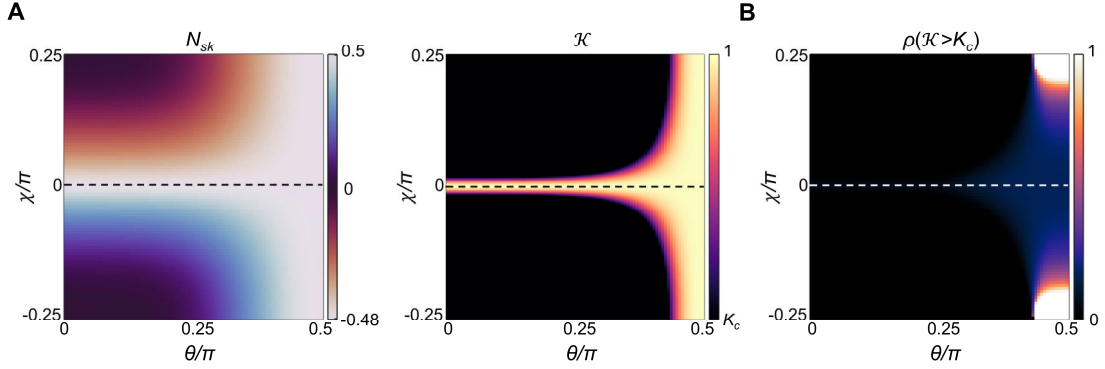

**Fig. S11. Evolution of ideal merons generated by elliptically polarized dipoles.** (**A**) Calculated skyrmion numbers $N_{sk}$ and $\mathcal{K}$ profiles under a $C_3$ configuration with $\xi = \pi/4$, and (**B**) Total density of states $\rho(\mathcal{K} > K_c)$ in the 3D polarization cube as functions of ellipticity $\chi$ and the deflection angle $\theta$. Here $K_c = 0.99$.

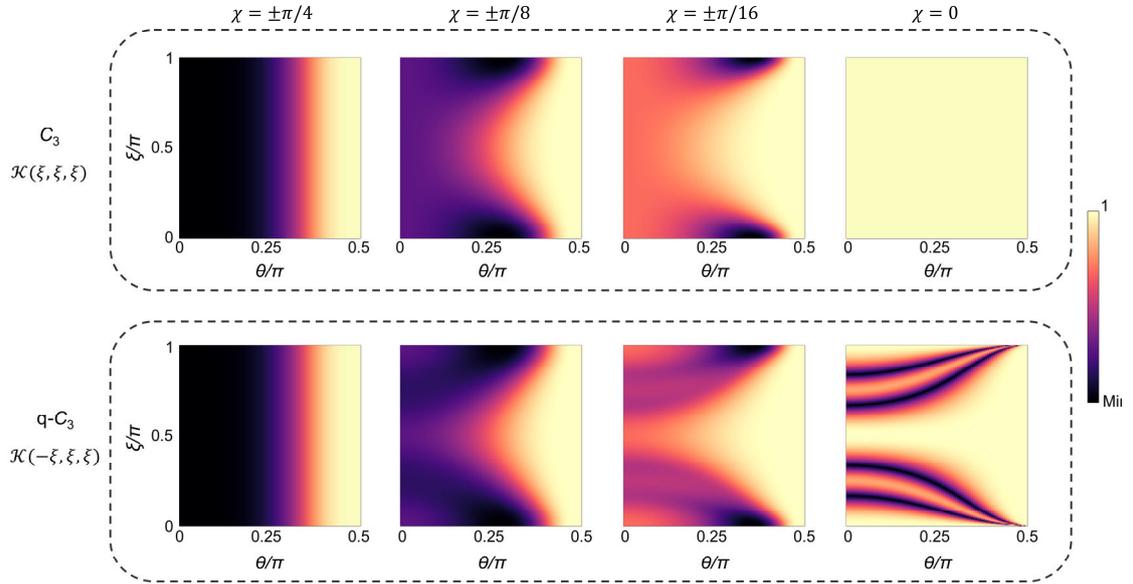

**Fig. S12. Influence of dipole ellipticity on the quasi-symmetry groups.** Calculated $\mathcal{K}(\pm\xi, \xi, \xi; \theta)$ distributions for the $C_3$ and q-$C_3$ groups across different ellipticity $\chi$, respectively.

## Section S6: Ubiquitous quasi-symmetry in general SOI systems

Beyond Rayleigh scattering (electric dipole), there also exists various non-paraxial optical systems, such as tightly focused beams (*11*, *12*) and Mie scattering (*13*),etc. Here, we generalize the formalism and existence conditions of quasi-symmetry groups in general SOIs of light.

S6.1 Strongly focused multi-beams.

As shown in Fig.S13A, the wavevector of the incoming collimated beam is rotated by a high numerical-aperture (NA) lens, generating a conical **k**-distribution in the focused region. As described by the Debye-Wolf theory (*14*), this refraction process is accompanied by adiabatic rotations of local polarization states which are attached and orthogonal to each **k**. Similarly, the focusing fields can be decomposed into a series of 3D rotations: $\hat{\mathcal{R}}_z(-\varphi)\hat{\mathcal{R}}_y(-\theta)\hat{\mathcal{R}}_z(\varphi)$. Explicitly, this transformation matrix in the circular basis takes the form:

$$\mathbf{E}(\mathbf{k}) = \sqrt{\cos\theta}\hat{V}^\dagger \hat{\mathcal{R}}_z(-\varphi)\hat{\mathcal{R}}_y(-\theta)\hat{\mathcal{R}}_z(\varphi)\hat{V}\mathbf{E}_i$$
$$= \sqrt{\cos\theta}\begin{pmatrix} A & -Be^{-2i\varphi} & \sqrt{2AB}e^{-i\varphi} \\ -Be^{2i\varphi} & A & \sqrt{2AB}e^{i\varphi} \\ -\sqrt{2AB}e^{i\varphi} & -\sqrt{2AB}e^{-i\varphi} & A-B \end{pmatrix}\mathbf{E}_i \quad (S30)$$

where $A = \cos^2(\theta/2), B = \sin^2(\theta/2)$ and $\sqrt{\cos\theta}$ is the apodization factor. It is evident that Eq.(S30) possesses a mathematical structure analogous to the dipole operator.

Consequently, we can construct topological structures through the interference of focused multi-beams by a high NA lens. We show it by generating the Stokes triangular meron lattices by focusing three collinearly polarized beams that possess threefold rotational symmetry about the *z*-axis, as shown in Fig.S13A and B.

For this multi-beam system, the commutation relation Eq. (3) still holds between the reduced SOI and mirror operator, ensuring that the previously established results regarding quasi-symmetry remain valid. Figure S13C shows that the evolution of the spin-maintained and flipped components $A$ and $B$ follow a different behavior compared to those in dipole radiation, yet they still converge to $A = B$ as $\theta \to \pi/2$. Hence, the density of polarized states still peaks at $\theta = \pi/2$ corresponding to maximal focusing aperture as shown in Fig.S13D. Moreover, we can straightforwardly generate ideal merons by tightly focusing the $C_3$ circularly polarized beams by a high-NA lens as described in Section S5.

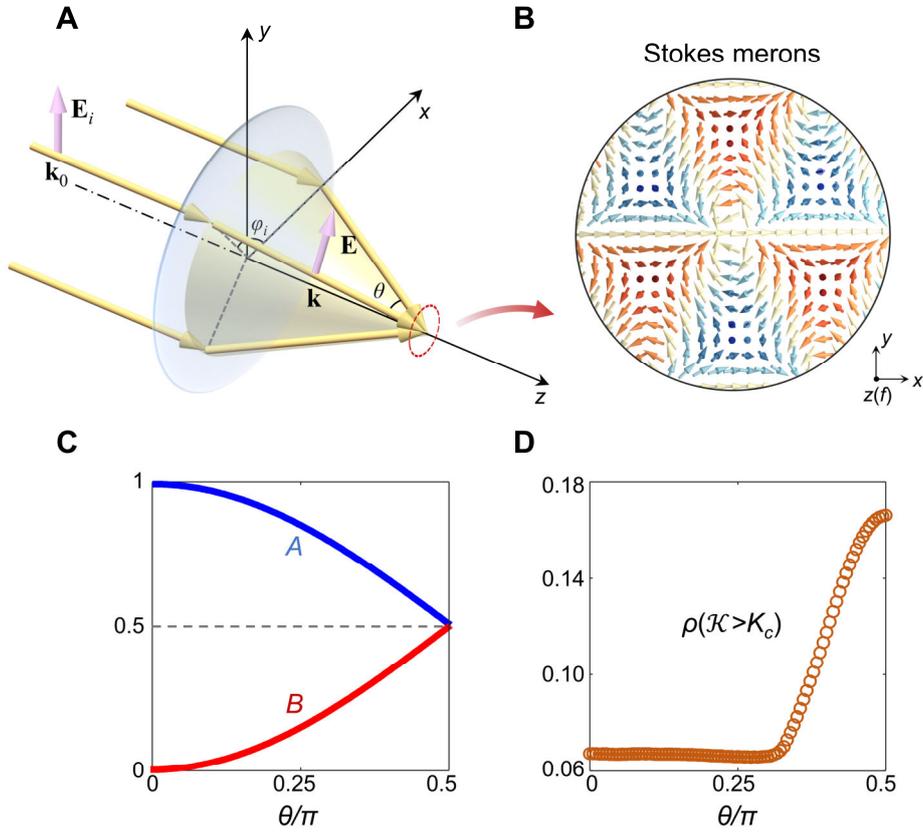

**Fig. S13. Stokes meron lattices formed by the interference of tightly focused $C_3$ beams via a high NA lens.** (**A**) Tight focusing of three paraxial polarized waves generates a perfect stokes meron lattice (**B**) on the focal plane. (**C**) SOI coefficients $A$ and $B$, and (**D**) total density of polarized states $\rho(\mathcal{K} > K_c)$, as functions of $\theta$. Here $K_c = 0.99$.

## S6.2 SOI operator for Mie scattering.

Mie theory exactly describes the optical absorption and scattering by small spherical particles of arbitrary radius and refractive index (*15*). The solution of Maxwell's equations in spherical geometry can be expressed as a superposition of vector spherical harmonics, which corresponds to the multipole expansion of the electromagnetic fields. Suppose an *x*-polarized plane wave propagating along the *z*-axis is incident on a spherical particle located at the origin, the scattered electric and magnetic fields take the form as following:

$$\begin{cases} \mathbf{E}_s = \sum_{n=1}^{\infty} E_n \left( i a_n \mathbf{N}_{e1n}^{(3)} - b_n \mathbf{M}_{o1n}^{(3)} \right) \\ \mathbf{H}_s = \frac{k}{\omega\mu} \sum_{n=1}^{\infty} E_n \left( i b_n \mathbf{N}_{o1n}^{(3)} + a_n \mathbf{M}_{e1n}^{(3)} \right) \end{cases}$$

where $E_n = i^n(2n+1)/n(n+1)$, $n \in \mathbb{Z}$. The functions $\mathbf{N}_{e,o1n}^{(3)}$ and $\mathbf{M}_{e,o1n}^{(3)}$ are vector spherical harmonics. The coefficients $a_n$ and $b_n$ represent the complex amplitudes of the electric and magnetic harmonic modes, which are determined by the incident wavelength and size/refractive index of the homogeneous particles.

Since we only focus on the propagating waves in the far-field region ($kr \gg 1$), where the scattered field is transverse ($\mathbf{k}_s \cdot \mathbf{E}_s \simeq 0$), the relation between incident and scattered field amplitudes is given (*15*):

$$\begin{pmatrix} E_{\|s} \\ E_{\perp s} \end{pmatrix} = \frac{e^{ik(r-z)}}{-ikr} \begin{pmatrix} S_2 & 0 \\ 0 & S_1 \end{pmatrix} \begin{pmatrix} E_{\|0} \\ E_{\perp 0} \end{pmatrix} \qquad (S31)$$

where

$$\begin{cases} S_1 = \sum_{n=1}^{\infty} \frac{2n+1}{n(n+1)} (a_n \pi_n + b_n \tau_n) \\ S_2 = \sum_{n=1}^{\infty} \frac{2n+1}{n(n+1)} (a_n \tau_n + b_n \pi_n) \end{cases} \qquad (S32)$$

Here the angle-dependent functions are defined as $\pi_n = P_n^1/\sin\theta$, $\tau_n = dP_n^1/d\theta$. ($P_n^1$ is the associated Legendre function). Equation (S31) bridges the field transformation between the global and local coordinate frames. We further develop it into a form analogous to Eq.(S3) via a series of pure geometric rotations as shown in Fig.S14, that is:

$$\mathbf{E}_s = \hat{V}^\dagger \hat{\mathcal{R}}_z(-\varphi)\hat{\mathcal{R}}_y(-\theta)\hat{M}\hat{\mathcal{R}}_y(\theta)\hat{\mathcal{R}}_z(\varphi)\hat{V}\mathbf{E}_0$$

$$\propto \begin{pmatrix} S_2\cos\theta + S_1 & (S_2\cos\theta - S_1)e^{-2i\varphi} & -\sqrt{2}S_2\sin\theta e^{-i\varphi} \\ (S_2\cos\theta - S_1)e^{2i\varphi} & S_2\cos\theta + S_1 & -\sqrt{2}S_2\sin\theta e^{i\varphi} \\ -\sqrt{2}S_2\sin\theta e^{i\varphi} & -\sqrt{2}S_2\sin\theta e^{-i\varphi} & 2S_2\tan\theta\sin\theta \end{pmatrix} \mathbf{E}_0 \quad (S33)$$

Here $\hat{M} = \mathrm{diag}(S_2/\cos\theta, S_1, 0)$ acts as a weighted projection operator in the local coordinate frame, modulating the projected field $\mathbf{E}_p = \mathbf{E}_0 - (\hat{e}_k \cdot \mathbf{E}_0)\hat{e}_k$ along the $\hat{e}_\theta$ and $\hat{e}_\varphi$ axes, respectively ($\hat{M}$ is not the mirror operator $\hat{\mathcal{M}}$). Considering the electric dipole approximation for particles whose size is much smaller than the incident wavelength, only the coefficient $a_1$ is non-zero. In this case, $\hat{M}$ reduces to the projector $\hat{\mathcal{P}}_z = \mathrm{diag}(1,1,0)$, and Eq.(S33) collapses to the form of Eq.(S3). It is noteworthy that the role of $\hat{M}$ is analogous to the planar anisotropic structures used in the geometric phase metasurfaces (*16*). However, this property of $\hat{M}$ does not originate from any structural anisotropy or inhomogeneities, but rather arises from coupling between multipole components.

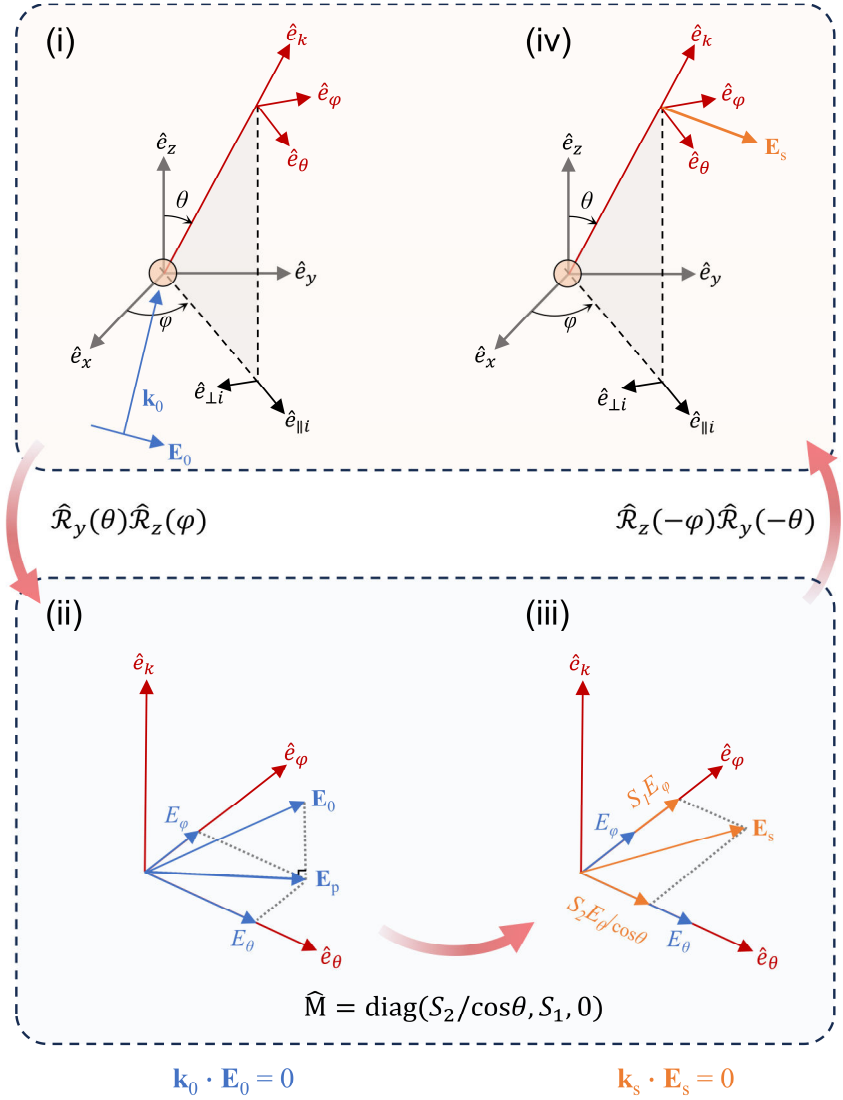

**Fig. S14. Schematic of the geometric transformation of the electric field in Mie scattering.** Several steps describing polarization transformation upon scattering: (i)–(ii) Successive rotations of the global Cartesian frame $(\hat{e}_x, \hat{e}_y, \hat{e}_z)$ to the local coordinate frame $(\hat{e}_\theta, \hat{e}_\varphi, \hat{e}_k)$: $\hat{\mathcal{R}}_y(\theta)\hat{\mathcal{R}}_z(\varphi)$. (ii)–(iii) describes the scattering process in the local coordinate frame, i.e., the anisotropic operation on the projected incident field. (iii)–(iv) Inverse rotations $\hat{\mathcal{R}}_z(-\varphi)\hat{\mathcal{R}}_y(-\theta)$ signify the transition back to the original global coordinate frame. Here $\hat{e}_\theta, \hat{e}_\varphi, \hat{e}_k$ are the orthonormal basis vectors associated with the radiation momentum of the scattered wave. Transverse condition is preserved for the incident (blue arrows) and scattered fields (orange arrows).

S6.3 General quasi-symmetry conditions.

Here we consider the $C_3$ Mie interference system analogous to the configuration described in Section S2, in which three linearly polarized monochromatic beams propagating along the $z$-axis excite Mie particles that possess identical electromagnetic responses. Based on Eq.(S33), we redefine the coefficients of the reduced SOI operator as follows:

$$\begin{cases} A = S_1 + S_2 \cos\theta \\ B = S_1 - S_2 \cos\theta \end{cases} \tag{S34}$$

where $A$ and $B$ can be complex if there is any non-zero or non-$\pi$ phase difference between any two different multipole components, so that $\mathcal{S}(\xi)$ is no longer guaranteed to be Hermitian. We calculate the allowed configurations for ideal $C_3$ lattices in the 3D parameter space ($\xi_1, \xi_2, \xi_3$) with varying $\theta \in (0, \pi/2]$ under several typical multipoles, and their combinations are shown in Fig.S15. For magnetic dipole ($b_1 \neq 0$), only the $C_3$ and anti-$C_3$ groups exist as four lines in 3D cube and remain stable for all values of $\theta$. Unlike the electric dipole, no proliferation of q-$C_3$ groups occurs. For electric quadrupole ($a_2 \neq 0$), the polarization configurations are strongly influenced by $\theta$, evolving continuously from four lines in paraxial regime to a bulky star-like region at $\theta = 0.25\pi$ corresponding to the emergence of q-$C_3$ groups, and eventually giving rise to three additional lines under grazing incidence that differ from the anti-$C_3$ group. For general combined multipoles(e.g., $a_1 = 1, a_2 = 1$), the occupancy and evolution of polarization states become more complex as varying SOIs.

Next, we discuss the existence conditions for a continuous quasi-symmetry group in Mie scattering. That is, at least one $\theta \in (0, \pi/2]$ exist to ensure $\mathcal{K}(-\xi, \xi, \xi) = 1$ for any $\xi \in [0, \pi)$. This requires that Eq.(S25) saturates the Cauchy-Schwarz inequality, yielding :

$$\begin{cases} |A| = |B| \neq 0 \\ \phi_{BA} = \arg(B) - \arg(A) = m\pi, m \in \mathbb{Z} \end{cases} \quad \exists\, \theta \in (0, \pi/2] \quad (S35)$$

Equation (S35) indicates that a critical SOI condition is necessary for the emergence of quasi-symmetry. As shown in Fig.S16A-C, the values of $\theta$ that satisfy $|A| = |B| \neq 0$ correspond exactly to the proliferation of q-$C_3$ groups. For electric dipole, $A = B \neq 0$ occurs at $\theta = \pi/2$. For magnetic dipole, $B$ is identically zero (i.e., no spin-flipped components), which implies the absence of quasi-symmetry groups. For the electric quadrupole, Eq.(S35) is achieved at $\theta = \pi/4$. Moreover, for general multipole combinations, q-$C_3$ groups can even emerge at several distinct $\theta$. Above results show that quasi-symmetry groups are universal and always correspond to the maximum occupancy of the polarization region, as shown in Fig.S16C. We can flexibly customize the optimal $\theta$ by controlling the size and refractive index of Mie particles.

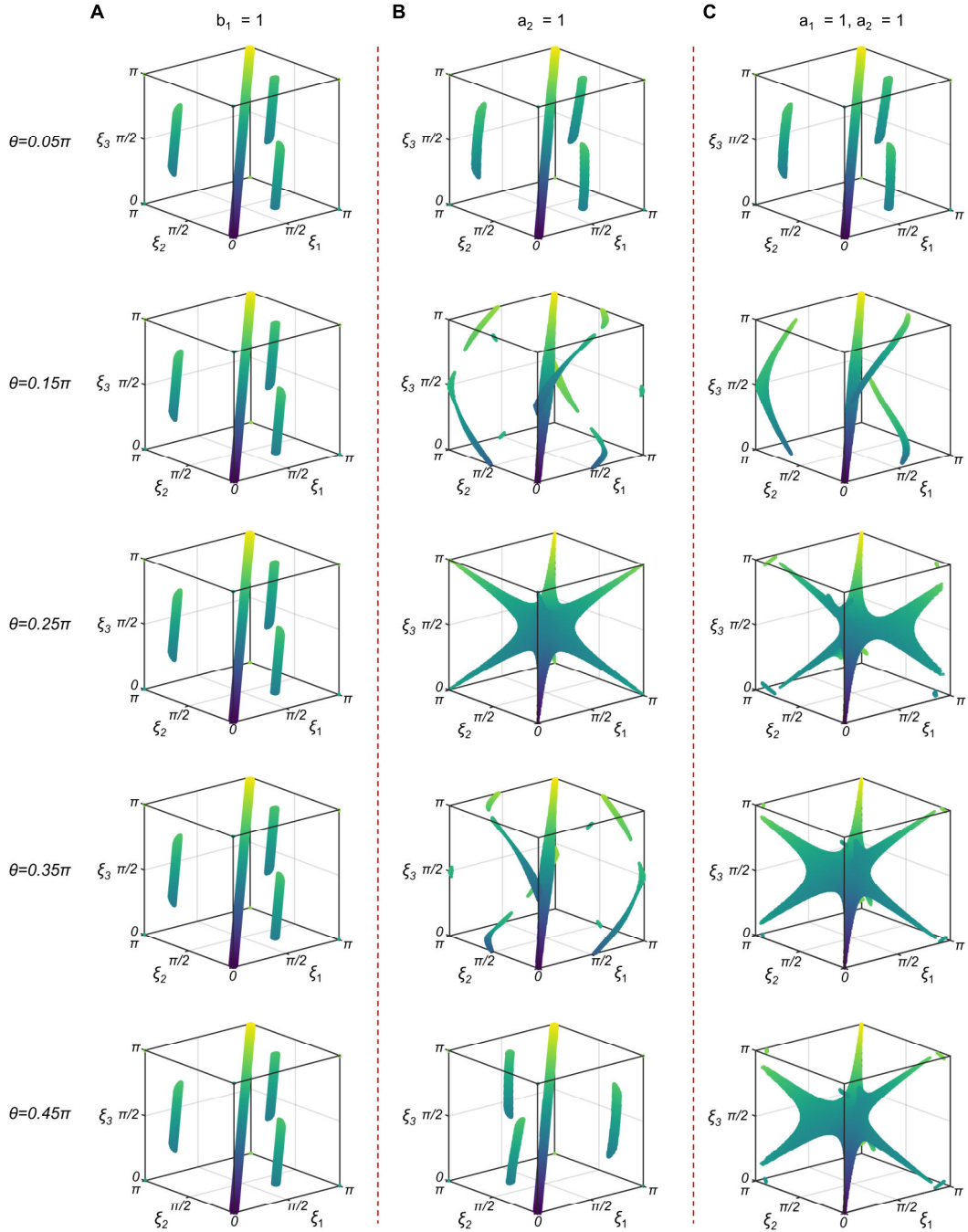

**Fig. S15. Evolution of polarization configurations with varying $\theta$ for different combinations of multipoles.** Polarization states in the 3D cube that form perfect $C_3$ lattices across distinct $\theta$ from top to bottom panels under different Mie responses: (**A**) magnetic dipole ($b_1 = 1$), (**B**) electric quadrupole ($a_2 = 1$), and (**C**) combination of electric dipole and quadrupolar combinations ($a_1 = 1, a_2 = 1$).

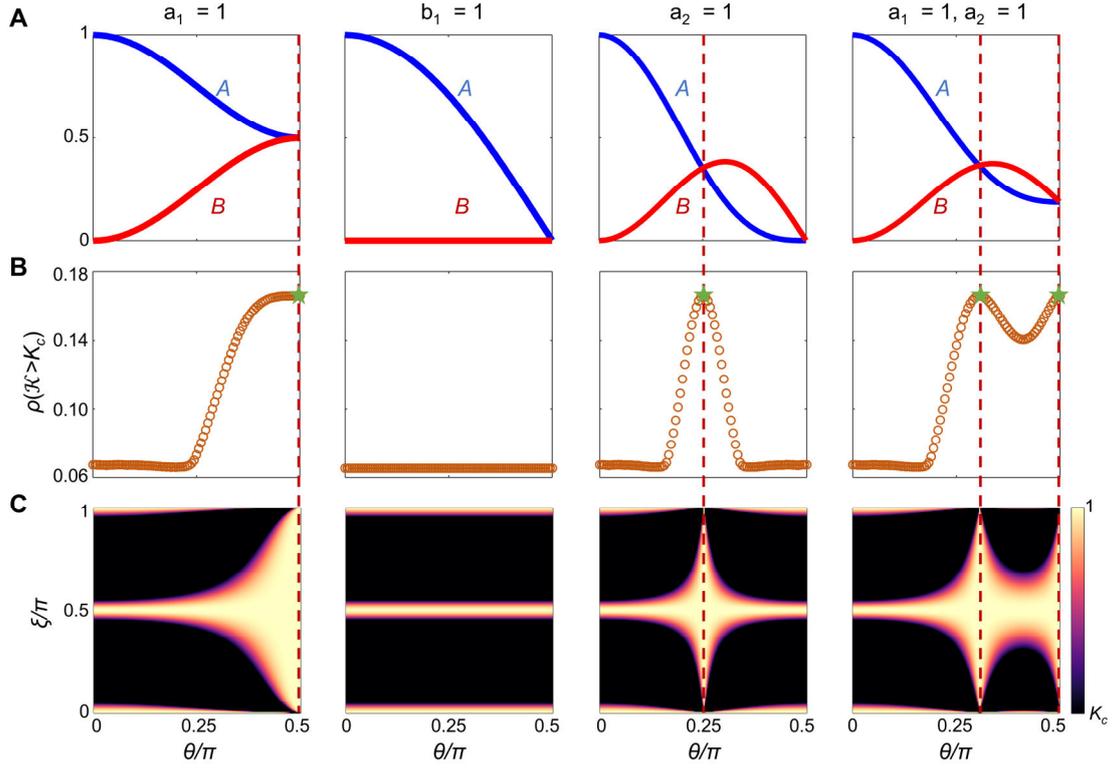

**Fig. S16. Conditions for the existence of quasi-symmetry groups under different Mie responses.** (**A**) Calculated SOI coefficients $A$ and $B$, (**B**) density of polarization states $\rho(\mathcal{K} > K_c)$, (**C**) $\mathcal{K}(-\xi, \xi, \xi)$ as functions of $\theta$ for different combinations of Mie coefficients. Red dashed lines mark the $|A| = |B| \neq 0$, and the green stars highlight the peak density of states. $K_c$ is set to 0.99.

# Section S7: Dipole Interference fields on an extended observation region

We show the distributions of the $C_3$ dipole interference field and the Stokes $S_3$ across an extended observation plane under a large deflection angle $\theta = 0.45\pi$, as shown in Fig.S17. High-quality $C_3$ meron lattices are generated in the central region of *xy* plane (Fig.S17A and B, insets), arising from the approximate preservation of the threefold rotational symmetry of the radiated momenta of dipoles about the *z*-axis. In addition, for circularly polarized illumination, a clear *spin-to-orbital angular momentum conversion* is observed in the vicinity of each dipole.

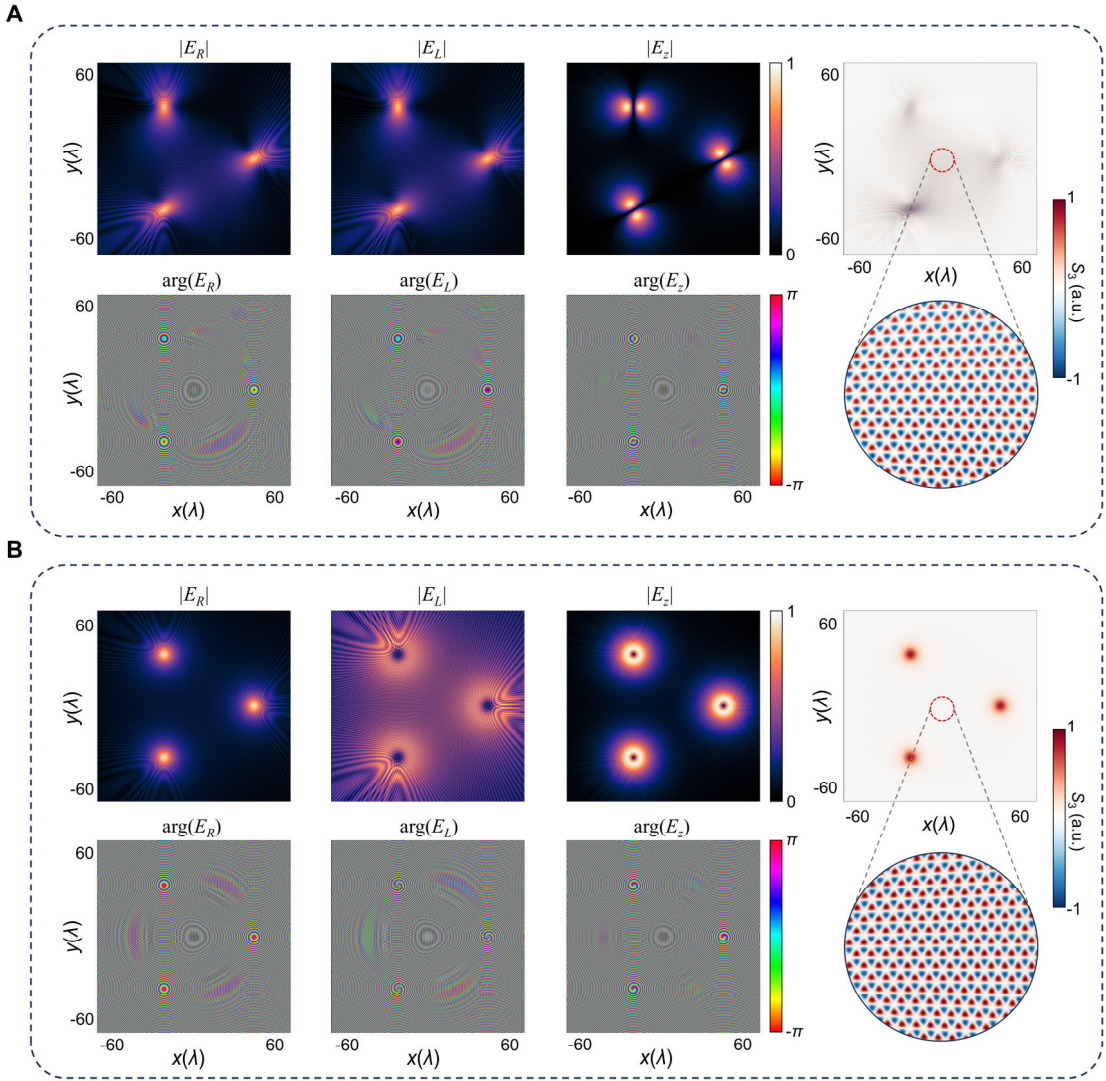

**Fig. S17. Distribution of interference field and Stokes $S_3$ across a large-scale observation region.** The circular polarization and longitudinal components,

$|E_R|, |E_L|, |E_z|$, the phase distributions of $\arg(E_R), \arg(E_L), \arg(E_z)$, and Stokes $S_3$ for (**A**) the q-$C_3$ dipoles configuration with $\xi = \pi/3$, and (**B**) three RCP dipoles at a deflection angle $\theta = 0.45\pi$. The radius of observation region is set to $40\lambda$.

## Section S8: A summary of various symmetry groups

Here, we summarize the various emergent polarization symmetry groups that arise in the construction of ideal topological lattices of light via multi-dipoles interference as shown in the following Fig.S18. The $C_3$ group exhibits the threefold rotational symmetry about the *z*-axis in both in-plane polarization states $\mathbf{E}_i$ and momenta $\mathbf{k}_i$. This group exhibits highly global symmetry due to the matching of $\mathbf{k}_i$ and $\mathbf{E}_i$, ensuring its survival during nonparaxial propagation (Fig.S18, middle panel). The anti-$C_3$ group emerges from an odd permutation of the polarization sequence of $C_3$ configuration and only survives under paraxial regime. It is inherently fragile due to the absence of global symmetry and vanishes completely under moderate SOI (Fig.S18, left panel). The q-$C_3$ group, defined by action of the partial mirror operation on any subsets of sources, exhibits broken symmetry under weak SOI. However, the in-plane field $\mathbf{E}(\mathbf{k}_i)$ along $\mathbf{k}_i$ is progressively reduced as enhanced SOIs (Highlighted arrows in right panel). At grazing incidence limit, any polarization vectors are completely perpendicular to corresponding $\mathbf{k}_i$, thereby facilitates the proliferation of the q-$C_3$ group.

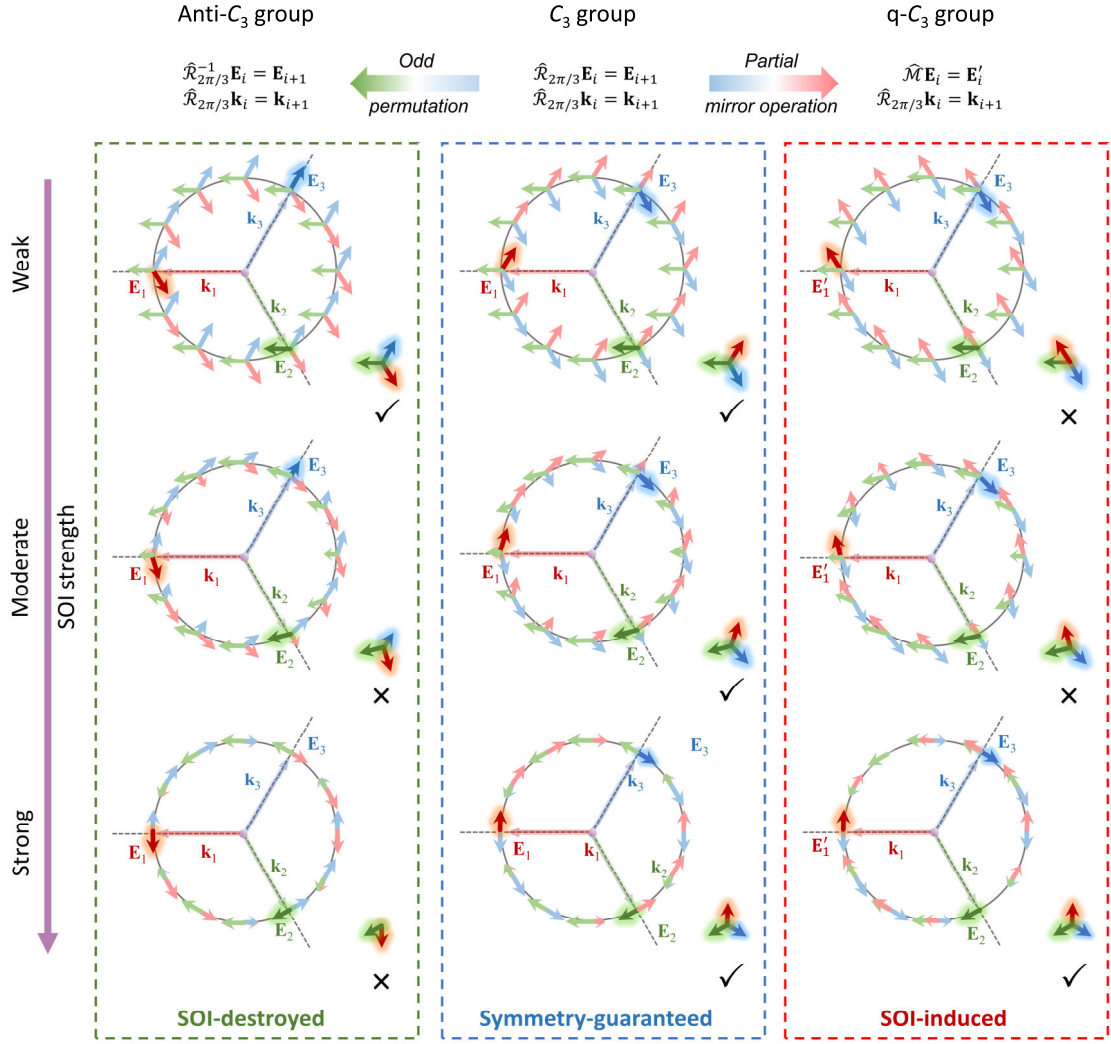

**Fig. S18. The properties of various polarization symmetry groups as a function of SOI.** From left to right panels, the defined anti-$C_3$, $C_3$ and q-$C_3$ groups arise from the distinct symmetry features of the in-plane polarization vectors $\mathbf{E}_i$ and corresponding radiation momenta $\mathbf{k}_i$ of the dipoles (Top panels). Evolution of the propagating **k**-cones and their attached polarization vectors (different-colored arrows) originating from different symmetry groups of sources under varying SOIs (Bottom panels).